\documentclass[preprints,article,accept,moreauthors,pdftex,10pt,a4paper]{mdpi}

\usepackage{amsmath, amsthm, amssymb, amsfonts, amsbsy,fixmath}
\usepackage{graphicx}
\usepackage{epstopdf}
\usepackage{upgreek}
\usepackage{booktabs}
\usepackage{multirow}
\usepackage{array}
\usepackage{float}
\usepackage{kantlipsum}
\usepackage{pifont}
\usepackage{caption}
\usepackage{hyperref}
\usepackage[ruled, lined, longend, linesnumbered]{algorithm2e}
\usepackage{algpseudocode}
\usepackage{tikz}
\usepackage{subcaption}
\usepackage{cleveref}
\usepackage{mathtools}

\firstpage{1} 
\pubvolume{xx}
\issuenum{1}
\articlenumber{5}
\pubyear{2018}
\copyrightyear{2018}
\history{Received: date; Accepted: date; Published: date}
\def\bf #1{\mathbf{#1}}
\def\cbr#1{\left\lbrace #1 \right\rbrace} 
\def\sbr#1{\left[ #1\right]} 
\def\nbr#1{\left( #1\right)}
\def\bs #1{\boldsymbol{#1}}
\def\B #1{\textbf{#1}}
\DeclareMathOperator*{\argmin}{argmin} 

\Title{   Dictionary Learning Phase Retrieval from Noisy Diffraction Patterns}

\Author{Joshin P. Krishnan $^{1,*}$, Jos\'e M. Bioucas-Dias $^{1}$ and Vladimir Katkovnik $^{2}$}

\AuthorNames{Firstname Lastname, Firstname Lastname and Firstname Lastname}

\address{%
	$^{1}$ \quad Instituto de Telecomunica\c c\~oes, Instituto Superior T\'ecnico, Universidade de Lisboa, Portugal; bioucas@lx.it.pt (J.M.B)\\
	$^{2}$ \quad Laboratory of Signal Processing, Technology University of Tampere, 33720 Tampere, Finland; vladimir.katkovnik@tut.fi (V.K.)}

\corres{Correspondence: joshin.krishnan@lx.it.pt (J.P.K)}

\abstract{This paper proposes a novel algorithm for image phase retrieval, i.e., for recovering complex-valued images  from  the amplitudes of noisy  linear combinations  (often the Fourier transform)  of the sought complex images. The algorithm is developed using the alternating projection framework and is aimed to obtain high performance for heavily noisy (Poissonian or Gaussian) observations. The  estimation of the target images is reformulated as a sparse regression, often termed sparse coding, in the complex domain. This is accomplished by learning a complex domain dictionary from the data it represents via matrix factorization with sparsity constraints on the code (i.e., the regression coefficients). Our algorithm, termed {dictionary learning phase retrieval} (DLPR), jointly  learns the  referred to dictionary and reconstructs the unknown target image.  The effectiveness of  DLPR is illustrated through experiments conducted on  complex   images, simulated and real,  where it shows noticeable advantages over the state-of-the-art competitors.}


\begin{document}
\section{Introduction \label{Introduction}}
\subsection{The Phase Retrieval Problem}
\textbf{Phase Retrieval} (PR) is an important and challenging problem in many fields of science and technology. PR is a crucial step in most diffraction- or scattering-based physical measurement systems. In such systems, the signals being used, namely a non-coherent photon flux, laser, X-ray, or any other variants of the electromagnetic radiation,  are diffracted through an object under  examination. These diffracted signals encode the structural information of the object such as thickness, density, or refractive index, and are detected using a suitable sensor. Since the detectors sense the diffracted signals by converting the photon flux to electrons, the information related to phase of the signal is not recorded. This is a major limitation as  important structural informations of the object are often coded mainly in the phase. PR aims at estimating the  hidden phase from the intensity measurements. This is a challenging problem as there is no one-to-one relation between the phase and the measured intensity, which is proportional to the magnitude square of the field.

This paper discusses the phase retrieval problem in an optical imaging scenario. We would like, however, to remark that, in spite of being developed in a specific scenario,  the proposed concept and algorithm is easily adapted to other imaging fields characterized by similar mathematical observation models. 

Hereafter, we consider that the power spectral density, i.e., the magnitude squared of the Fourier transform, is the measurement provided by the sensor.  This assumption is very reasonable and common in literature \citep{2013_candes_phaselift,2017_Katkovnik_Phase_Retrieval,2015_Candes_PRWF}. It is based on the fact that the optical wavefront at the sensor plane in the far field (i.e., at a large enough distance from the imaging plane) is well approximated by the power spectral density of the wavefront at the object plane. This result can easily be derived from the Fraunhofer diffraction equation for coherent imaging systems \citep{2007_Saleh_Fundamentals, goodman2005intro}.

\begin{figure}[h!]
	\centering
	\includegraphics[width=0.8\textwidth]{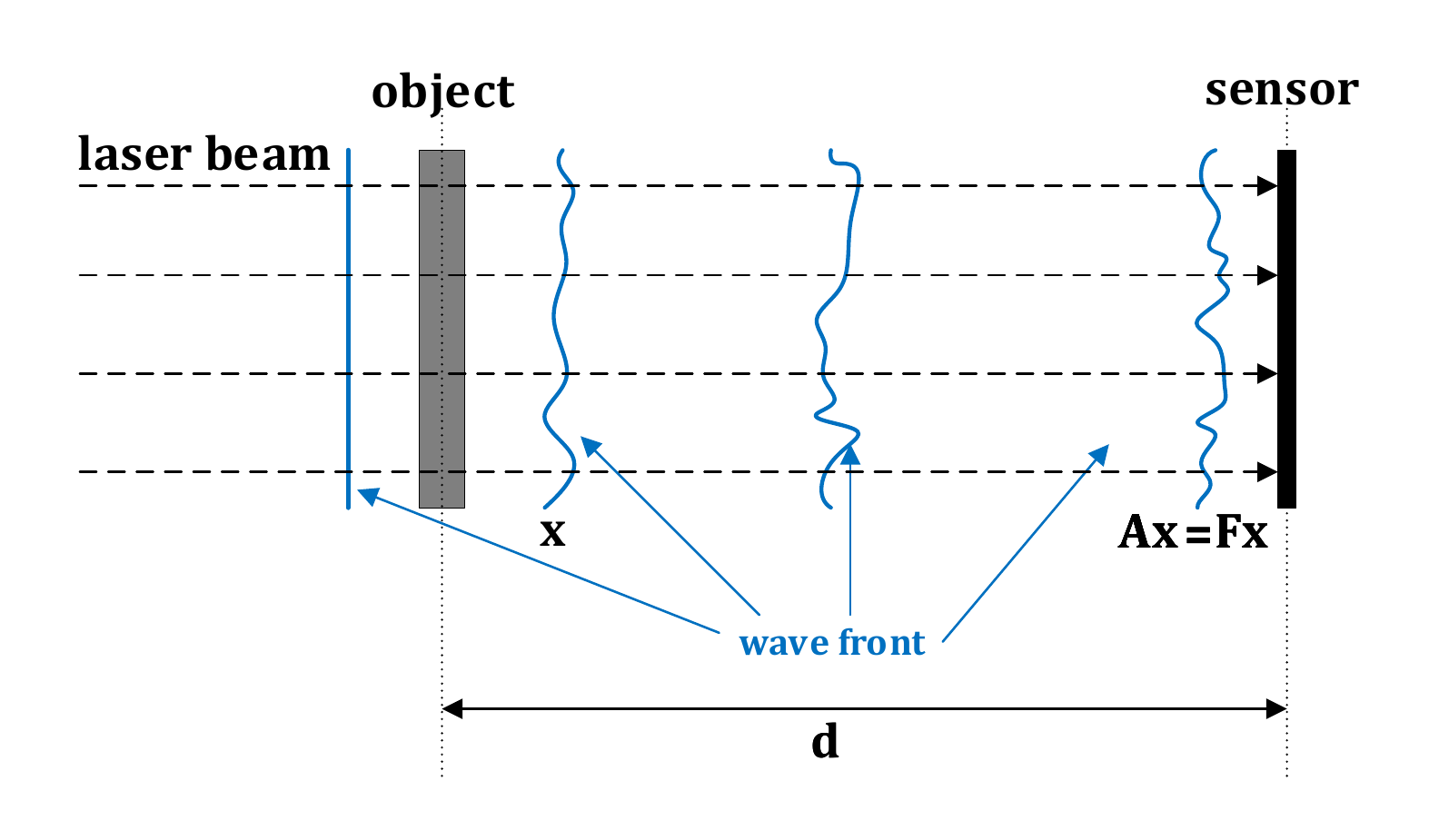}
	\caption{Optical setup of a lensless imaging system.}
	\label{setup_basic}
\end{figure}

Fig. \ref{setup_basic}, courtesy of \cite{2017_Katkovnik_Phase_Retrieval}, schematizes a lensless optical imaging system. A planar wavefont produced by a laser beam \footnote{Here, without loss of generality, we assume that the intensity of the laser beam is constant over the object.} is transmitted through an object (whose image to be inferred) and propagates until it reaches the sensor. The 2D PR problem in the vectorized form is as follows:
\begin{align}
&\textrm{find} &\bf{x} & \in \mathbb{C}^n \nonumber \\
&\textrm{subject to} &\bf{z}&={|\bf{A}\bf{x}|}^2 + {\bf w}\in
\mathbb{R}^n,\hspace{2cm}&
\label{PReqn}
\end{align}
where $\bf{x}$ is the complex-valued wavefront with $n$ pixels at the object plane and $\bf{A} \in \mathbb{C}^{n\times n}$ is an $n\times n$ matrix modeling the wavefront propagation from the object to the sensor plane, which, in the far field, is well approximated by DFT matrix $\bf{F}\in \mathbb{C}^{n\times n}$, i.e., $\bf{A}=\bf{F}$,  $|\cdot|$ is the component-wise magnitude operator, and $\bf w$ accounts for measurement (or model) noise. The presence of the term $|{\bf A \bf x}|$ in \eqref{PReqn} makes this problem nonconvex and thus challenging.  
 
\subsection{Phase Retrieval: Applications, algorithms \& recent trends}
Phase retrieval field has a rich history, perhaps the proposal of phase contrast imaging in 1930 by Frits Zernike (Nobel prize 1953) is one of its early stage milestones. Another revolutionary contribution was in 1952 by  David Sayre in the field of crystallography \cite{1952_Sayre_Some}. Perhaps the most important field that is benefited from PR algorithms is X-ray crystallography \cite{1990_Millane_PR_in_cryst}, \cite{1993_Harrison_Phase_prob_cry}. X-ray phase-contrast imaging (XPCI) is a powerful tool in structural investigations \cite{1965_Bonse_XRAY,2003_Petrakov_Xray,1995_Snigirev_possib,1996_Wilkins_Phase,2006_WPfeiffer_Phase}. As already mentioned before, most of the physical measurement systems record only the intensity of the detected signal and this makes PR an unavoidable process in various fields, namely microscopy \cite{2008_Miao_Extending}, optics \cite{1963_Walther_Question_of_PR}, speech and acoustics \cite{1993_Rabiner_fund_speech_rec}, \cite{2006_Balan_On_signal}, astronomical imaging \cite{1987_Dainty_Image_Recovery}, array imaging \cite{2011_Chai_Array}, interferometry \cite{2017_Demanet_Convex}, computational biology \cite{1978_Stefik_Inferring}, diffraction imaging \cite{2007_Bunk_Diffractive}, blind deconvolution \cite{2004_Baykal_Blind}, quantum mechanics \cite{2006_Corbett_pauli}, \cite{1965_Reichenbach_Philosophic}, quantum information \cite{2013_Heinosaari_Quantum}, blind channel estimation in wireless communications  \cite{2014_Ahmed_Blind} \cite{Ranieri_phaseretrieval}, X-ray tomography \cite{2010_Dierolf_Ptychographic}, differential geometry \cite{2002_Bianchi_solution} etc.

The early approaches to PR  fall under two categories of methods: error-reduction  and gradient-based. The phase-retrieval problem  defined in \eqref{PReqn} is often formulated as the following least squares problem or empirical risk ($\mathcal{R}({\bf x}) :=||\bf{z}-{|\bf{A}\bf{x}|}^2 ||^2$)  minimization problem:
\begin{equation}
\min_{\bf{x}\in\mathbb{C}^n} ||\bf{z}-{|\bf{A}\bf{x}|}^2 ||^2. \label{LSpblm}
\end{equation}
\textit{Error-reduction} algorithms are iterative and  guarantee to reduce error ($\mathcal{R}$) in each iteration. The most popular family of PR algorithm, the Gerchberg-Saxton (GS) \cite{1972_GS_orig} and its variants \cite{2015_Guo_GS}, \cite{1994_Yang_GSYangGU}, comes under this category. These are iterative algorithms based on alternating projections between the object plane and the diffraction (Fourier) plane. A basic GS algorithm has four simple steps: (1) Forward projection: Fourier transform of the object wavefront; (2) imposing Fourier magnitude constraints at the Fourier plane: replace the modulus of the forward propagated wavefront with the measured intensity to form an estimate of the Fourier transform; (3) backward projection: inverse Fourier transform operation; and (4) imposing spatial-domain constraints at the object plane: modifying the backward projected wavefront by imposing support constraints in accordance with the prior information of the object.

Fienup, in his famous Hybrid Input-Output (HIO) algorithm \cite{1982_Fienup_comparison}, addressed the slow convergence of the GS algorithms by imposing additional spatial-domain corrections. Although both GS, HIO and their variant algorithms are widely used in optical imaging, they suffer from stagnation at  local minima due to the non-convex nature of the Fourier magnitude constrains. 

An alternative class of methods to address PR is  based on \textit{gradient searches}. In this class, in each iteration, the wavefronts are updated in such a way that the partial derivative of the gradient of an error metric, which one seeks to minimize (e.g., $\mathcal{R}$), is equated to zero. This line of attack is discussed in \cite{1982_Fienup_comparison}, where, although the convergence rate of a steepest descent-based \cite{1982_Fienup_comparison} algorithm is low, its conjugate gradient-based version \cite{1991_Lane_CG} is much faster.

The recent Wirtingling Flow (WF) algorithm \cite{2015_Candes_PRWF} is an iterative complex domain gradient descent technique. Although the gradient operation is not well-defined for complex domain variables, WF adopts a surrogate derivative, termed the ``Wirtinger derivative'',  and is characterized by  special features, such as spectral initialization, non-trivial step-size parameter, etc. The truncated version of the WF algorithm, termed truncated Wirtingling Flow (TWF) algorithm \cite{2015_Chen_TWF} improves the  initialization and descent procedures in an adaptive fashion by a statistically motivated regularizing technique, which filters out the terms bearing too much influence on the initial estimate or search directions. A surprising claim stated in \cite{2015_Chen_TWF}  is  that solving the quadratic equations \eqref{PReqn}  is ``nearly as easy as solving linear equations'', which is  a significant leap in many fields.

An alternative phase retrieval strategy has been proposed recently by exploiting the semi-definite relaxations. The set of quadratic equations represented by \eqref{PReqn} is rewritten as linear equations in a higher dimension space. The phaseLift \cite{2013_candes_phaselift} and PhaseCut \cite{2015_Waldspurger_phasecut} are two important algorithms in this category. In PhaseLift, the ``lifting'' is done by the variable transformation $\bf{X}:=\bf{xx}^H$,\footnote{$H$ is the Hermitian operator} which provides a convenient linear constraint in terms of the matrix variable $\bf{X}$. The underlying rank minimization non-convex problem is relaxed to  a convex one yielding a semidefinite program (SDP). Similarly, in PhaseCut, the complex variable $\bf{x}$ is separated into an amplitude component and a phase component, and only the phase component is `lifted' and then optimized via SDP. But the matrix lifting in SDP-based approaches results a higher-dimensional variable, which, in turn, makes the algorithm computationally demanding compared to the alternating projection approaches.

A computationally light PR algorithm based on convex relaxation that operates in the natural domain of the signal is proposed in \cite{2017_bahmani_convex}. In this work, a ``complex polytope'' of feasible solutions is considered by relaxing the quadratic equations of phaseless measurements to inequalities. The desired solution would be one of the extreme points of this polytope which is found using a convex program.

In the recent decades, PR has been highly benefited from the advances in the field of compressed sensing and new imaging technologies. In line with the overview \cite{2015_PRoverview_Yonina}, we wish to mention two important approaches in phase retrieval which have been sprouted as a results of these advancements. They are (1) Phase front modulation based approaches and (2) Sparsity based approaches.

\subsubsection{Phase front modulation}
\label{wfrnt_modifn}
Phase front modifications \cite{2003_Nugent_wavefrnt_modifcn}, \cite{2008_Johnson_mask} are  imaging techniques in which a known phase modulation is intentionally introduced into the object field. 
These phase modification are done through phase masks mounted on the wavefront propagation path and different masks lead to different diffraction patterns. An analogy from microscopy would be focusing one part of the sample to be examined and then repositioning the sample to exploit the spatial diversity. By shifting the phase mask or, equivalently, by using different masks, independent diffraction patterns, termed  coded diffraction patterns (CDPs)\cite{2015_CANDES_PR_cdp}, are generated from the same illumination area of the object. The CDP introduces over-determination in the amplitude attenuation and phase shift of the wave diffracted through the object. Also, in a CDP, the recoded intensity at a sensor point  have contributions from all points in the object. Spreading the information of one object point to all pixels of the sensor minimizes the the sensor noise and also mitigates the stagnation in the retrieval process \cite{Zhang_2007_aper_plane_modifcn}. 

\begin{figure}[h!]
	\centering
	\includegraphics[width=0.8\textwidth]{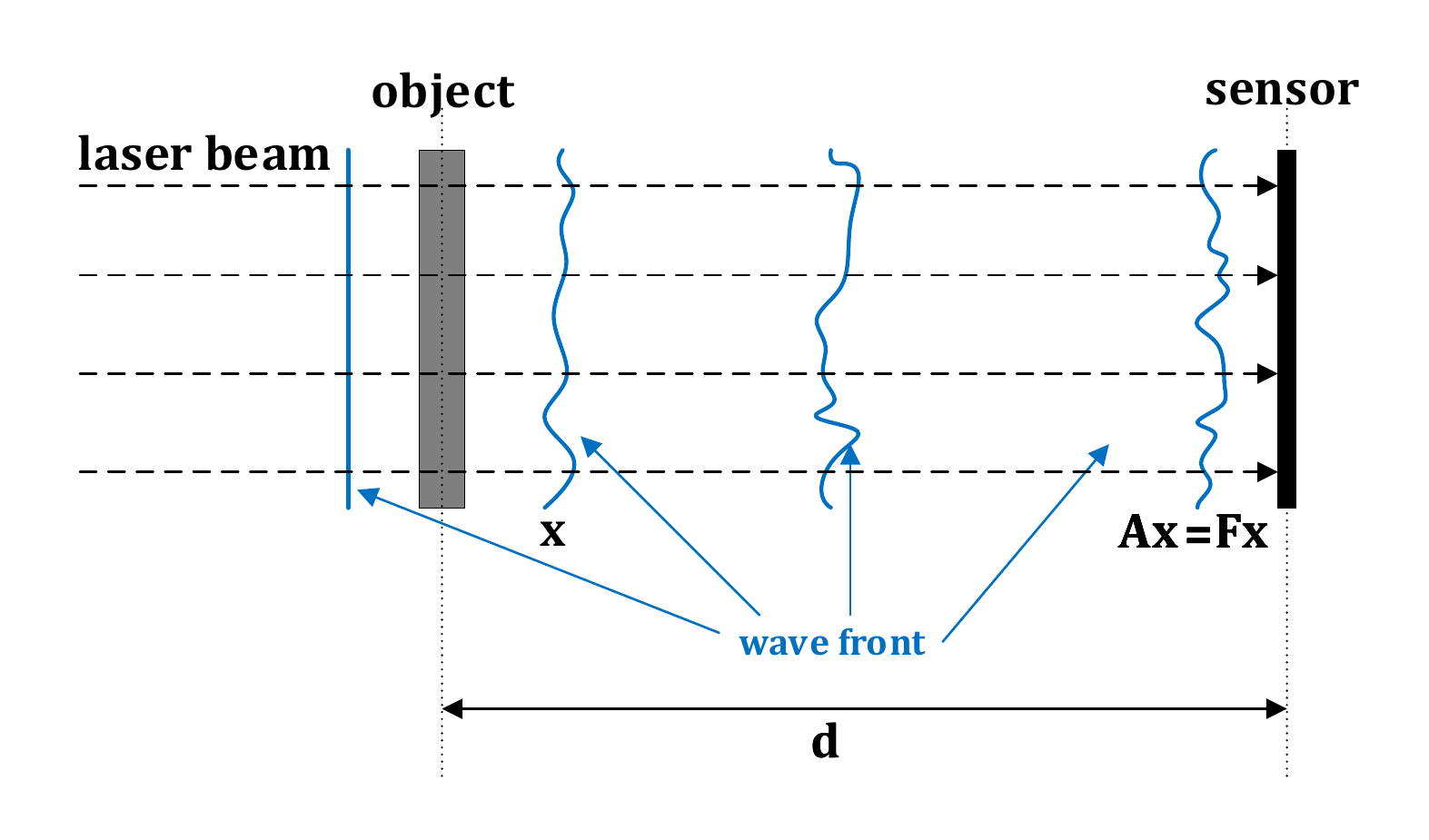}
	\caption{Optical setup for a lensless imaging system using a phase mask.}
	\label{setup_mask}
\end{figure}

The setup shown in Fig. \ref{setup_mask} corresponds to a modifications of that of Fig. \ref{setup_basic},
where a phase mask is incorporated. The corresponding PR problem is 
\begin{align}
&\textrm{find}       &\bf{x}~& \in\mathbb{C}^n \nonumber \\
&\textrm{subject to} &\bf{z}_s&   ={|\bf{A}_s\bf{x}|}^2+{\bf w}\in\mathbb{R}^n, ~~~s=1, 2, 3....S  \label{PReqn_mask},
\end{align}
where  $\cbr{\bf{A}_s}_{s=1}^{s=S}\in\mathbb{C}^{n\times n}$ are the wavefront propagation matrices given by $\bf{A}_s=\bf{FM}_s$. In our model,  $\bf{F}\in \mathbb{C}^{n\times n}$ denotes the DFT matrix  and $\bf{M}_s\in\mathbb{C}^{n\times n}$ is a phase mask which is a diagonal matrix of complex exponents, i.e., $\bf{M}_s=\text{diag}\cbr{e^{j\bs{\phi}_1^s},~e^{j\bs{\phi}_2^s}, \dots, e^{j\bs{\phi}_n^s}},~j=\sqrt{-1}$, where $\bs{\phi}_i^s,~i=1,2,\dots,n$, are random phase values.
\subsubsection{Sparsity meets Phase Retrieval}
\label{sparsity_subsec}
The concept of sparsity has received tremendous attention from various signal processing areas. Signals and images of the real world admit sparse representations when represented on suitable frames \cite{2010_Elad_sparse}, which can be exploited to build low dimensional models. As in many areas, sparsity has become a hot topic in PR methods and algorithms. TSPR \cite{2017_Jaganathan_TSPR} and GESPAR \cite{2014_Shechtman_GESPAR} are two important state-of-the-art algorithm that exploit sparsity. TSPR  is a two-stage algorithms in which the support of the signal is estimated in the first stage and the signal is estimated in the second stage using a sparsity constrained PhaseLift framework along with the learned support. GESPAR  uses an optimization-based greedy algorithm in which PR is reformulated as a sparsity-constrained least square problem.

Another important research line in the context of sparsity exploits the  self-similarity \cite{2006_Dabov_bm3d,2015_Teodoro,2006_elad_image} exhibited by the natural images, which has been extensively exploited via patch-based approaches in various fields of image processing \cite{2007_Dabov_Image,2014_Li_CT_sparse,2011_Deledalle_NL}. Since the phase images are natural images, they often exhibit high level of self-similarity. A number of phase imaging algorithms exploiting self-similarity has been recently introduced, namely the complex dictionary-based SpInPhase \cite{2015_Hongxing_Interferometric}, the complex domain Mixture of Gaussian -based MoGInPhase \cite{2017_Joshin_MoGInpahse},   the BM3D-based CBM3D\cite{2017_KATKOVNIK_CBM3D}, and the sparse approximation of the object phase and amplitude SPAR \cite{2017_Katkovnik_Phase_Retrieval}.

SPAR \cite{2017_Katkovnik_Phase_Retrieval} is a recent PR algorithm that shows remarkable performance in highly noisy scenarios, compared to the state-of-the-art algorithms. SPAR is built on a classical GS framework with additional features of random phase modulation of the wavefront and sel-similarity  regularization of the phase and amplitude. The two stages of filtering, i.e.,  filtering  the noisy (Poissonian) observations at the sensor plane and  filtering the phase and the amplitude at the object plane, make this algorithm a step forward in noisy PR. SPAR exploits the self-similarity of complex domain patches by separately applying sparsity to the phase and amplitude components. Since the phase is restored modulo$-2\pi$, the recovering of the absolute phase is carried out by including a phase unwrapping \cite{2007_Bioucas_Phase} step. This is a sensitive part of SPAR, as the phase unwrapping is  known to be an  NP-hard problem. Another downside of SPAR is  its  high computational complexity, compared with that of GS. 

\subsection{Proposed algorithm and Contribution}
\label{contri}
Inspired by SPAR \cite{2017_Katkovnik_Phase_Retrieval} and by SpInPhase \cite{2015_Hongxing_Interferometric},  we propose a new algorithm, termed Dictionary Learning Phase Retrieval (DLPR), for PR from noisy diffraction pattern. DLPR is derived under a variational framework and is developed for both Poissonian and Gaussian observation models. We adopt the classic alternating projection-based framework (GS \cite{1972_GS_orig}), and incorporate 1) the wavefront modification detailed in Section {\ref{wfrnt_modifn}}, 2) filtering at the sensor plane for Poissonian (or Gaussian) observation, and 3) filtering at the object plane using dictionary-based sparse coding in the complex domain. We remark that, recently, a dictionary based PR algorithm is proposed in \cite{2016_Tillmann_DOLPHIn}, which makes use of a real valued dictionary and deals with real valued object, whereas DLPR deals with complex valued dictionary and object, which is an essential requirement in a practical PR scenario that often involves complex valued wavefronts. One of the remarkable advantages of sparse modeling in the complex domain is that it gets rid of the ambiguity due to phase wrapping, which makes it very robust to the heavily noisy observations, compared to the SPAR algorithm. The main contributions of the proposed work is summarized below:
\begin{enumerate}
	\item A variational reformulation of the PR problem that incorporates a dictionary based sparse regression in the complex domain.
	
	\item An algorithm that jointly retrieves phase and learns the dictionary yielding sparse representations (codes) for the complex domain patches of the object wavefront.
	
	\item An extension of the algorithm to a class-specific scenario, where the dictionary is learned from clean images  of the same class.
\end{enumerate} 

The paper is organized as follows: in Section \ref{Problem-formulation}, the sparsity modeling for the complex domain wavefront is discussed. This section also introduces Poissonian and Gaussian observation models. The DLPR algorithm is derived in Section \ref{algorithm}, by solving  a step-by-step variational formulation for the forward propagation, the sensor plane filters for Poissonian and Gaussian observations, the backward propagation, and the sparse modeling at the object plane. In Section \ref{SecExp}, an experimental  study and characterization of the algorithm performance are provided including comparisons with  relevant state-of-the-art algorithms.

\section{Problem formulation \label{Problem-formulation}}
\subsection{Sparse regression-based wavefront modeling \label{Sparse}}
\begin{figure}[h!]
	\begin{subfigure}[width=\textwidth]{.49\linewidth}
		\centering
		\includegraphics[scale=0.49]{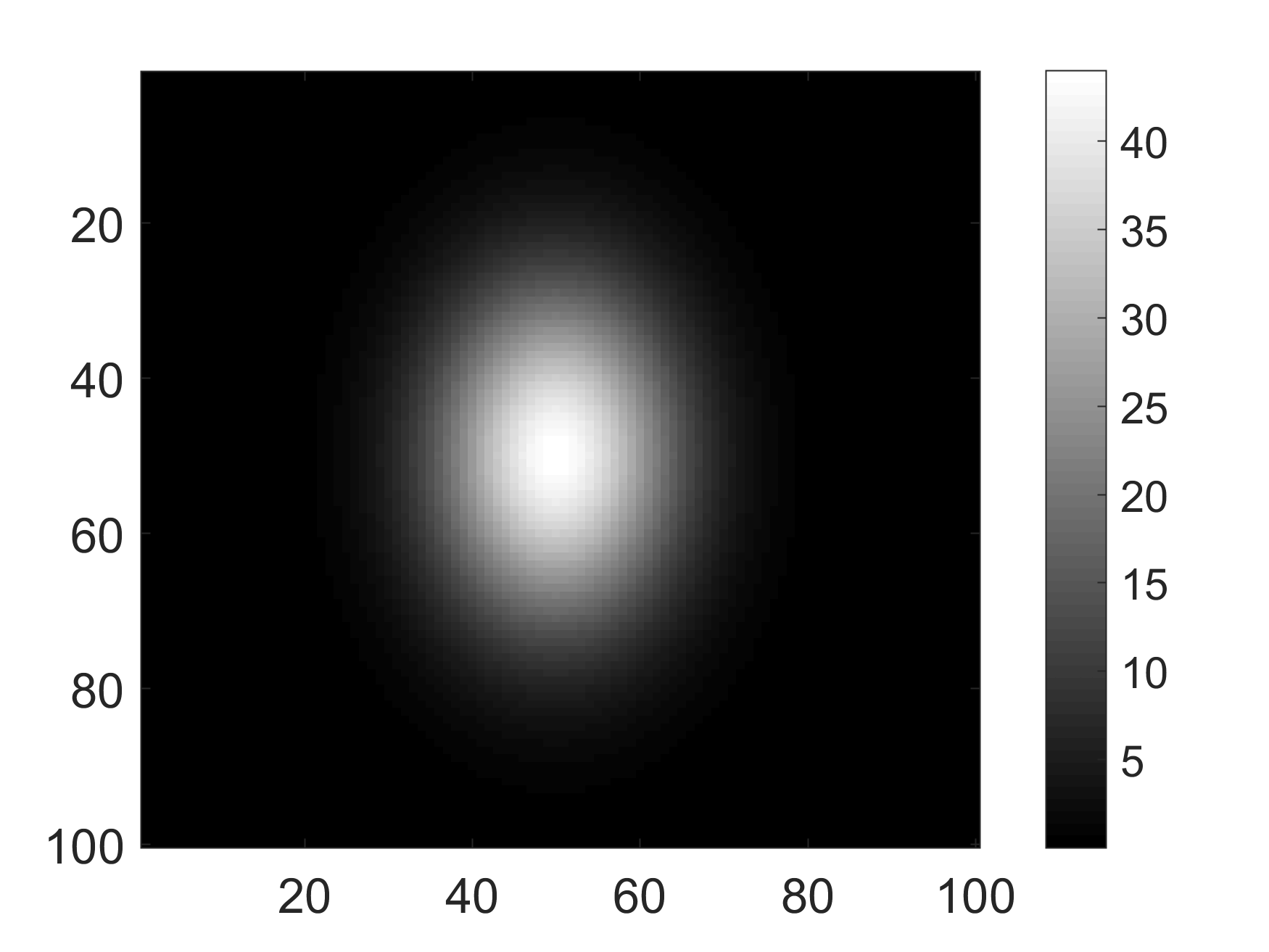}
		\caption{Absolute phase:$\bs \psi$.}
	\end{subfigure}
	\begin{subfigure}[width=\textwidth]{.49\linewidth}
		\centering
		\includegraphics[scale=0.49]{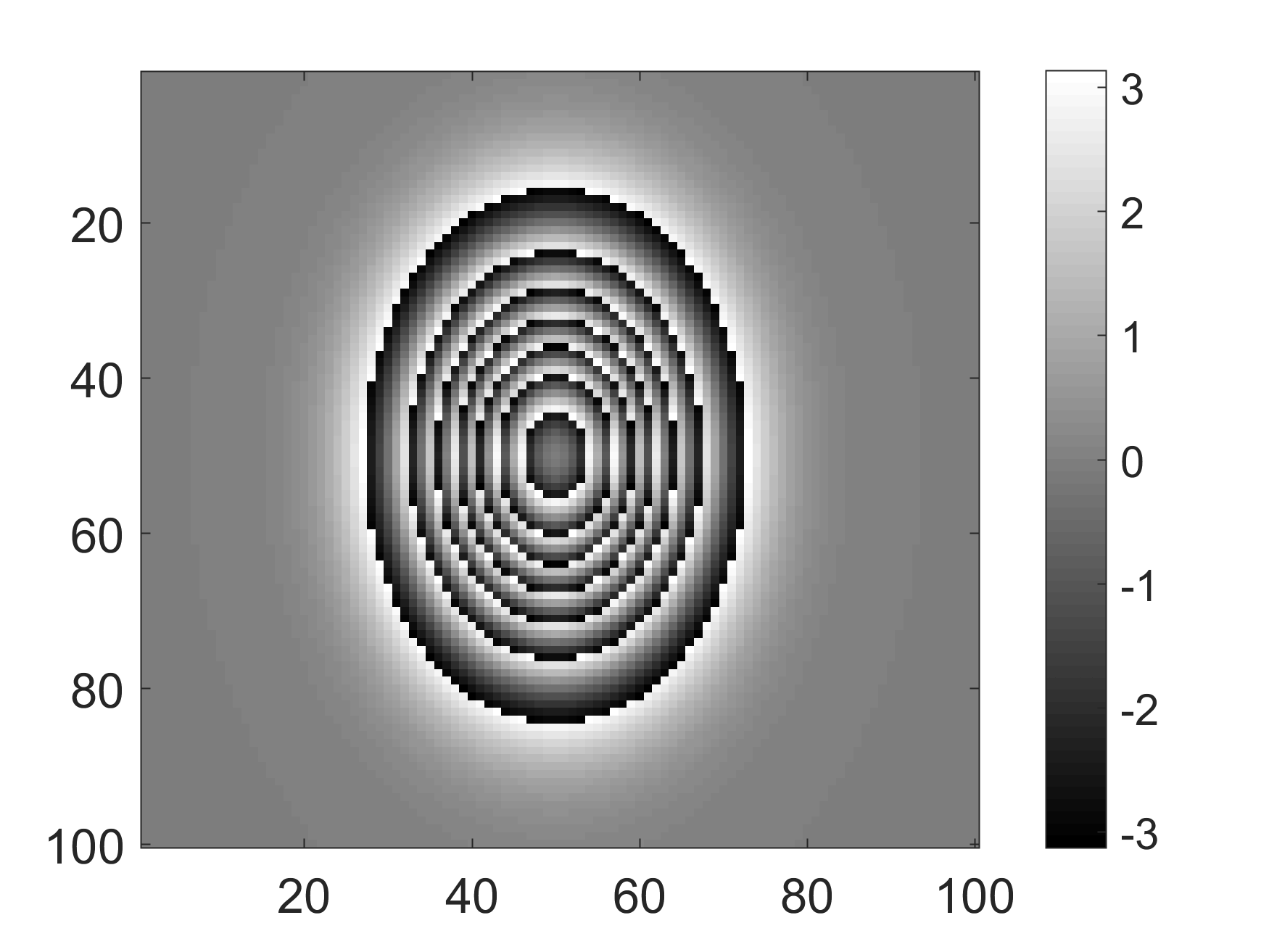}
		\caption{Interferometric phase:$\bs \psi _{2\pi}$.}
	\end{subfigure}  
	\caption{Illustration of Phase wrapping using a Gaussian surface.}
	\label{Pwrap}   
\end{figure}

Let the vectorized complex domain image  wavefront $\bf{x} \in \mathbb{C}^n$ be represented as
\begin{align}
\bf{x}:=\bf{a\odot} e^{j\bs \psi}, \label{ceqn}
\end{align}
where $\bf{a} \in \mathbb{R}_+^n$ is the positive amplitude, $\bs{\psi} \in \mathbb{R}^n$ is the absolute phase of the object wavefront and the operation $\bf{\odot}$ stands for the element-wise (Hadamard) multiplication. Herein, all functions applied to vectors are to be understood as component-wise. One important factor to be noted here is that the accessible phase of a complex wavefront, which is extracted from the $2\pi$-periodic complex sinusoidal function, undergoes phase wrapping as defined below:
\begin{eqnarray}
\mathcal{W}:\mathbb{R} &\rightarrow & \left[ -\pi, \pi\right) \nonumber \\ 
\bs{\psi} & \mapsto & \mod(\bs{\psi}+\pi, 2\pi)-\pi, \label{eqn:wrap}
\end{eqnarray}
where $\mathcal{W}$ is the wrapping operator that performs the  2$\pi$-modulo phase wrapping operation. In the ensuing text, we use the notation $\bs {\psi}_{2\pi}:= \mathcal{W}(\bs {\psi})$, and term $\bs {\psi}_{2\pi}$ as the \textit{interferometric phase}. We remark that the interferometric phase may be directly obtained from the image $\bf x$ as  $\bs {\psi}_{2\pi}:= \text{angle}({\bf x})$.
The interferometric phase is a non-linear function of the absolute phase $\bs \psi$ and possess a pattern like discontinuities, called \textit{interferometric fringe patterns},  as illustrated in Fig. \ref{Pwrap}.

Regarding the  sparse modeling of a set of images ${\bf x}\in\mathbb{C}^n$, one can  consider  their amplitude  and   phase  independently. However, from $\bf x$, or any estimate of it, we have only access to the corresponding interferometric phase, which raises  a number of issues on how to model that phase, since the sparse models  are usually specified in terms of the absolute phase.   A solution to address this issue is to perform  phase unwrapping using a suitable algorithm (e.g., PUMA \cite{2007_Bioucas_Phase}), obtaining estimates of the absolute phase and then performing  sparse modeling on it, as done in \cite{2017_Katkovnik_Phase_Retrieval}. But the unwrapping algorithms are highly sensitive to noise and image discontinuities larger than $\pi$. Also, in many practical applications, the phase and amplitude are strongly correlated and a decoupled sparsification of them  is equivalent to assume that they are statistically independent, thus failing to exploit the correlation that might exit between them.

Here, we  adopt the viewpoint introduced in \cite{2015_Hongxing_Interferometric} to introduce a new PR strategy in which sparse modeling of the wavefront is done in the complex domain. This new strategy has the following distinct advantages:  1) inherent exploitation of the phase-amplitude correlation, as they are treated together in the complex domain form of wavefront, and 2) increased robustness to the noise, as the phase unwrapping is not used.  

Sparse and redundant representations are  recent and active research topics in signal and image processing \cite{2010_Elad_sparse} and have wide applications in many areas, such as denoising, restoration, interpolation, compression, sampling, recognition etc., to list only a few. The patch-based approaches are very popular in this context. Motivated by the recent phase imaging techniques \cite{2015_Hongxing_Interferometric}, \cite{2017_Joshin_MoGInpahse}, \cite{2014_Katkovnik_Phase}, \cite{2017_KATKOVNIK_CBM3D}, we stress that, if $\bf a$ and $\bs \psi $ are self-similar, then their complex version $\bf{a\odot} e^{j\bs \psi}$ should also be self-similar (see \cite{2015_Hongxing_Interferometric} for further details). This means that it is possible to find similar  complex domain patches located at different parts of the wavefront, which opens the door to sparse representations, where each complex image patch is represented by a linear combination of a small number of vectors, often termed \textit{atoms}, from a (possibly learned) dictionary. Herein, we adopt this perspective to the optical imaging scenario of interest. The assumption of self-similarity results naturally from the fact that the involved amplitude and phase images are from real world objects. 

As a natural follow up of the above discussion, we put forward a new idea for an optical wavefront modeling based on sparse regression using complex domain dictionaries. We follow a patch-based approach for the sparse wavefront modeling. Let ${\bf R}_i\in \{0,1\}^{w^2\times n}$ and ${\bf x}_i^p := {\bf R}_i\bf{x} \in \mathbb{C}^{w^2}$ denote, respectively, a matrix that extracts the $i$ square patch of size $w\times w$ from the wavefront image $\bf x$  and the extracted patch of size $w^2$. It is to be noted that the superscript $p$ in ${\bf x}_i^p$ indicates that ${\bf x}_i^p$ is a patch and the same notation is used hereafter. We also, remark that the rows of ${\bf R}_i$ are a subset of the rows of the identity matrix of size $n$. The index $i\in {\cal I}_p$ of a given  patch  refers to the wavefront image pixel corresponding to the top left hand side of the patch. Using these definitions, the patch aggregation, i.e., patches to image transformation can be realized using the following equation:
\begin{align}
\bf{x}=\nbr{\sum_{i\in{\cal I}_p}\bf{R}_{i}^{H} \bf{R}_{i}}^{-1}\nbr{\sum_{i\in{\cal I}_p}\bf{R}_{i}^{H}{\bf{x}}_i^p} \label{p2imag},
\end{align}
where $H$ is the Hermitian operator and $\sum_{i\in{\cal I}_p}\bf{R}_{i}^{H}{\bf{x}}_i^p$ is the sum of all patches ${\bf{x}}_i^p$ put back into their original locations.  We emphasis that the matrix  $\sum_{i\in{\cal I}_p}\bf{R}_{i}^{H} \bf{R}_{i}$ is diagonal and its $j^{th}$ diagonal element indicates the number of times the pixel $j$ appears in the set of extracted patches. The patches are extracted and later aggregated in an overlapping manner and we refer to \cite{2015_Hongxing_Interferometric} for more details on the decomposition and composition of the patches.

To better understand the sparse regression, at this point, we assume that a dictionary ${\bf D} \equiv \sbr{\bf{d}_1, \bf{d}_2, ..., \bf{d}_k}\in \mathbb{C}^{w^2\times k}$ for sparsely representing a complex patches ${\bf x}_i^p$, $i\in{\cal I}_p$, is available. By sparse regression, we mean that any patch ${\bf x}_i^p$, $i\in{\cal I}_p$, may be represented as a linear combination of few columns, often termed \textit{atoms}, from $\bf{D}$. Regarding the patch ${\bf x}_i^p$,  our sparse wavefront modeling is formally defined as follows:
\begin{align}
\widehat{\bs {\alpha}}_i&=\argmin_{\bs{\alpha}} ||\bs{\alpha} ||_0 \text{ s.t.:} ~~ ||\bf{x}_i^p-\bf{D}\bs{\alpha} ||^2 \leq \delta,   \nonumber\\
\widehat{\bf{x}}_i^p&=\bf{D}\widehat{\bs {\alpha}}_i, \label{SpReg}
\end{align}
where $\bs {\alpha} \in \mathbb{C}^k$ is the optimization \textit{code} for $\bf{x}_i^p$, $\delta \geq 0 $ is a parameter controlling the reconstruction error, and $||\bs \alpha||_0$ denotes the $\ell_0$ pesudo-norm of $\bs {\alpha}$,  i.e., the number of nonzero elements of the vector $\bs \alpha$. The constrained optimization \eqref{SpReg} has the following interpretation: since we are minimizing the zero norm of the code, the optimization aims to find an $\bs{\alpha}$ with the  fewest number of non zero entries, keeping the reconstruction error $||\bf{x}_i^p-\bf{D}\bs{\alpha} ||^2$ below a pre-controlled value $\delta$. We incorporate the sparse regression  \eqref{SpReg} in the phase retrieval algorithm to model the object wavefront and this, in turn, serves as a filter at the object plane of the optical setup. The algorithms for learning  dictionary $\bf{D}$ and code $\bs{\alpha}$ will be discussed later in the Section \ref{algorithm}.

\subsection{Noisy observation modeling \label{P_Observations}}
The observation model adopted in many PR algorithms assumes a noiseless scenario. But in practice, the measured intensities at the sensors are often corrupted by noise. Two common sources of noise are 1) the random motion of the charge carrier such as thermal noise generated in the sensor, which is modeled by  additive white Gaussian perturbations, and 2) photon-limited measurements, which are modeled by  Poissonian perturbations. So, we develop the DLPR algorithm by considering either Poissonian or Gaussian noise observation models.

\subsubsection{Poissonian observation model}
The optical imaging systems usually capture the wavefront intensity as the counts of photons hitting a detector. There are many practical limitations that restrict us from using a high energy radiation with sufficient number of photons. For instance, in medical X-ray imaging, a high energy radiation can cause damage to the specimen, which is often a human organ. Also, an imaging system that can deploy only low exposure time or can use only limited amount of light (e.g., night vision and astronomical imaging systems) results in a low  number of photons. Phase imaging using such low energy radiation is termed as photon-limited phase imaging, which results in heavily noisy registered measurements. The widely used additive Gaussian noise model fails to accurately model such discrete observations and, instead, a Poisson distribution shall be used \cite{2016_Soulez_Proximity}, \cite{2010_Raginsky_Compressed}, \cite{2014_Salmon_PCA},\cite{2010_Harmany_Phot_lmtd_imagng}.

In the presence of Poissonian noise, we assume that the components  $\bf{z}_s[l]$, for $l=1,\dots,n$ and $s=1,\dots,S$, have Poisson distributions with parameters $ {\bf y}_s[l] : = |{\bf A}_s{\bf x}|^2[l]$; that is
\begin{align}
\bf{z}_s[l] \sim {\cal P} \nbr{\bf{y}_s[l]} \label{nf_pois},
\end{align}
where ${\cal P}(\theta)$ stands for Poisson distribution with parameter $\theta \geq 0$.
Now let us consider a photon-limited imaging scenario and use the scaling factor $\chi >0$ to scale  the mean value of the photon count. We restate \eqref{nf_pois} as
\begin{align}
\bf{z}_s[l] \sim {\cal P} \nbr{\bf{y}_s[l]\chi} \label{eqnoisy_pois}.
\end{align}
Here $\chi$ is an important parameter that controls the photon-count, which in turn controls the noise level. In a practical imaging system, $\chi$ is a function of various factors such as exposure time, sensor sensitivity, etc. Using the Poisson distribution formula, model \eqref{eqnoisy_pois} can be rewritten as
\begin{equation}
p(\mathbf{z}_{s}[l]=k)=\exp (-\mathbf{y}_{s}[l]\chi )\frac{(\mathbf{y}%
	_{s}[l]\chi )^{k}}{k!}\text{,}  \label{obs-3}
\end{equation}%
where $k\in \mathbb{N}_0$ is the photon count at the sensor. To understand the effect of $\chi$ on the noise level, we define signal-to-noise ratio (SNR) at the sensor as the ratio between the square of the mean and the variance (equal to mean for Poisson random variable) of $\mathbf{z}_{s}[l]$, yielding
\begin{align}
\textrm{SNR}:=\frac{\mathbb{E}^{2}[\mathbf{z}_{s}[l]]}{\mathbb{V}[\mathbf{z}_{s}[l]]}
=\mathbf{y}_{s}[l] \chi. \label{eqnSNR1}
\end{align}
Expression  \eqref{eqnSNR1} shows that SNR grows linearly with $\chi$. Definition \eqref{eqnSNR1} motivates us to write  a global estimate of SNR,  as
\begin{align}
\textrm{SNR}_\textrm{global}:=\frac{\sum_{s=1}^{S} \sum_{l=1}^{n} \mathbb{E}^{2}[\mathbf{z}_{s}[l]]}{\sum_{s=1}^{S} \sum_{l=1}^{n} \mathbb{V}[\mathbf{z}_{s}[l]]} 
= \frac{\sum_{s=1}^{S} \sum_{l=1}^{n}\mathbf{y}_{s}[l]^2 \chi}{\sum_{s=1}^{S} \sum_{l=1}^{n}\mathbf{y}_{s}[l]}\label{eqnSNR2}
\end{align}
\subsubsection{Gaussian observation model\label{G_Observations}}
We also consider the widely used Gaussian noise that models, e.g.,  the thermal noise at the sensor. In this case, the observation model is
\begin{equation}
\mathbf{z}_{s}[l]=\mathbf{y}_{s}[l]+\sigma \bs{\varepsilon }_{s}[l],\quad l=1,\dots,n,\; s=1,\dots,S,
\label{GaussianNoise}
\end{equation}%
where ${\bf y}_{s}[l] = \mathbb{E}[{\bf z}_s[l]] = |{\bf A}_s\bf x|^2[l] $,  $\bs{\varepsilon }_{s}[l]\sim \mathcal{N}(0,1)$ is a set of independent and identically distributed (i.i.d.) random variables, and $\sigma $ stands for the standard deviation of the noise.
\section{Dictionary Learning Phase Retrieval (DLPR) algorithm }
\label{algorithm}
We propose a novel idea, which we term as DLPR, to address the PR problem  discussed in \eqref{PReqn_mask}. DLPR is an alternating minimization algorithm that exploits the dictionary based sparse modeling of the object wavefront. The sparsification improves the conditioning of the inverse problem and its robustness to noise. The algorithm is derived for both Poissonian and Gaussian observation models and is developed based on the optical setup as discussed in Fig. \ref{setup_mask}.

\subsection{DLPR for Poissonian Observation model} 
\label{sec:Poisdervn}
Referring to the Poissonian observation model \eqref{obs-3}, we state our PR problem as the estimation of $\bf{x}\in \mathbb{C}^n$ from the noisy observations $\{\mathbf{z}_{s}\}_{s=1}^{s=S}$.  The corresponding negative log-likelihood (after neglecting the constant terms) is
\begin{equation}
\mathcal{L}({\bf x})=\sum_{s=1}^{S}\sum_{l=1}^{n}[|\bf{A}_{s}\bf{x}|^{2}[l]\chi -\bf{z}_{s}[l]\log (|\bf{A}_{s}\bf{x}|^{2}[l]\chi )],
\label{Likelihood_a}
\end{equation}
or equivalently
\begin{align}
\label{Likelihood}
\mathcal{L}(\{\bf{u}_{s}\},\bf{x})& =\sum_{s=1}^{S}\sum_{l=1}^{n}[|\bf{u}_{s}[l]|^{2}\chi -\bf{z}_{s}[l]\log (|\bf{u}_{s}[l]|^{2}\chi )] \\
\nonumber
\text{s.t.:} ~~~\bf{u}_s &=\bf{A}_{s}\bf{x},\quad s=1,\dots,S.
\end{align}
A  likelihood formulation similar to \eqref{Likelihood} is used in \cite{2015_Candes_PRWF} and \cite{2015_Chen_TWF} in which the focus is to maximize $\mathcal{L}(\{\bf{u}_{s},\bf{x}\})$ by computing  $\partial \mathcal{L}(\{\bf{u}_{s}, \bf{x}\})/\partial \bf{x}$ using Wirtinger derivatives \cite{2011_Adali_complex_valued} in an iterative manner. Here, however, we take a different approach by converting  the hard constraint into a quadratic penalty and including a dictionary-based sparse regression term, which promotes image self-similarity. The DLPR variational formulation is as follows:
\begin{align}
(\{\widehat{\bf{u}}_{s}\}, \widehat{\bf{x}}, \{\widehat{\bs{\alpha}}_i\},\widehat{\bf{D}}) 
= & \underset{\{{\bf{u}}_{s}\}, {\bf{x}}, \{{\bs{\alpha}_i}\},{\bf{D}}}{\arg\min}
\mathcal{L}(\{\bf{u}_{s}\}\text{, }\bf{x}\text{,}\{\bs{\alpha}_i\}\text{, }\bf{D}) \nonumber \\
& \text{s.t.:} ~~{\bf D}\in {\cal C},
\end{align}
with
\begin{align}
\label{DLPR}
&\mathcal{L}(\{\bf{u}_{s}\}\text{, }\bf{x}\text{,}\{\bs{\alpha}_i\}\text{, }\bf{D})\nonumber \\  & =\sum_{s=1}^{S}\sum_{l=1}^{n}[|\bf{u}_{s}[l]|^{2}\chi -\bf{z}_{s}[l]\log (|\bf{u}_{s}[l]|^{2}\chi )] +\frac{1}{\gamma}||\bf{u}_{s}-\bf{A}_{s}\bf{x}||_{2}^{2} +\sum_{i\in{\cal I}_p}\tau_{a}||\bs{\alpha}_i||_{0} + \beta||\bf{R}_i\bf{x}-\bf{D}{\bs \alpha }_i||_{2}^{2},
\end{align}
where $\cal C$ is a convex set to be defined later, $\{{\bf{u}}_{s}\}:= 
\{ \bf{u}_{s}\}_{s=1}^{s=S}$, and $\{\bs{\alpha}_i\}:= \{\bs{\alpha}_i\}_ {i\in{\cal I}_p}$. The parameter $\tau_{a}\ge 0$ controls the level of sparsity and $\gamma,\beta >0$ control the weights of the quadratic penalties.
Let us introduce the  notation
\begin{align}
g_{s}(\bf{u}_{s}[l])& :=[|\bf{u}_{s}[l]|^{2}\chi -\bf{z}_{s}[l]\log (|\bf{u}_{s}[l]|^{2}\chi )]\text{.} \\
\bf{v}_{s} & :=\bf{A}_{s}\bf{x}.,  \label{DLPR2}
\end{align}
and use it to rewrite  \eqref{DLPR} as 
\begin{align}
\mathcal{L}(\{\bf{u}_{s}\}\text{, }\bf{x}\text{, }\{\bs{\alpha}_i\}\text{, }\bf{D}) =\sum_{s=1}^{S}\sum_{l=1}^{n}g_{s}(\bf{u}_{s}[l])+ \frac{1}{\gamma}\sum_{s=1}^{S}||\bf{u}_{s}-\bf{v}_{s}||_{2}^{2} 
+\sum_{i\in{\cal I}_p}\tau_{a}||\bs{\alpha}_i||_{0}+\beta||\bf{R}_i\bf{x}-\bf{D}{\bs \alpha }_i||_{2}^{2}. \label{DLPR3}
\end{align}

The objective function in \eqref{DLPR3} is non-convex w.r.t. to the variables $(\{\bf{u}_{s}\}\text{, }\bf{x}\text{, }\{\bs{\alpha}_i\},\bf{D}$).  We address this issue by alternating minimization methods, which is a common approach to tackle this sort of non-convexity \cite{2013_Rakotomamonjy_AltMin}. As per this strategy, the likelihood  $\mathcal{L}$ is optimized by considering one variable at a time, treating the others as constants. Below we derive the optimization w.r.t. each variable. For each optimization, we rewrite the objective function by considering the terms depending on the optimization variable and disregarding the other terms.
\subsubsection*{\textbf{ Problem 1: Optimization w.r.t. $\{\bf{u}_{s}\}$}}  
\vspace{-0.8cm}
\begin{align}
\bf{\widehat{u}}_{s}=\arg \min_{\{\bf{u}_{s}\}} \sum_{s=1}^{S}\sum_{l=1}^{n}g_{s}(\bf{u}_{s}[l]) + \frac{1}{\gamma}\sum_{s=1}^{S}||\bf{u}_{s}-\bf{v}_{s}||_{2}^{2}.\label{UsMin}
\end{align}
Given that \eqref{UsMin} is decoupled w.r.t. $\bf{u}_s[l]$, for $s=1,\dots,S$ and $l=1,\dots,n$, we 
have $\widehat{\bf u}_s=(\widehat{\bf u}_s[1],\dots, \widehat{\bf u}_s[n])$, where $\widehat{\bf u}_s[l]$ is the proximity operator \cite{2011_combettes_proximal} of the function $g_s \gamma/2$ computed in  \cite{2016_Soulez_Proximity} and given by (see \cite{2014_Katkovnik_Wavefront} for further details)
\begin{align}
\bf{u}_{s}[l]&=\bf{b}_{s}[l]\exp (j\, \text{angle}(\bf{v}_{s}[l]\text{)}), \label{Stage_1_2}\\
\shortintertext{where} 
\bf{b}_{s}[l]&=\frac{|\bf{v}_{s}[l]|+\sqrt{|\bf{v}_{s}[l]|^{2}+4\bf{z}_{s}[l]\gamma(1+\gamma \chi )}}{2(1+\gamma\chi )} \text{.}  \label{bslPoi}\\
\shortintertext{For large values of the scale factor $\chi$, we  have}
\lim_{\chi \to\infty} \bf{b}_{s}[l]&=\sqrt{\mathbf{z}_{s}[l]/\chi }.\label{lim1}
\end{align}
Also from \eqref{obs-3}, it follows that a very high value of the scaling factor $\chi$ models the noiseless observations where the Poissonian model can be replaced with a deterministic model with probability 1. In this case we can write
\begin{align}
\lim_{\chi \to\infty} \mathbf{z}_{s}[l]/\chi&=\mathbf{y}_{s}[l].\label{lim2}\\
\shortintertext{ Using \eqref{lim1} and \eqref{lim2}, we rewrite \eqref{Stage_1_2} to obtain the amplitude update formula, for noiseless observation, as follows:}
\mathbf{u}_{s}[l]&=\sqrt{\mathbf{y}_{s}[l]}\exp (j\, \text{angle}\,(\mathbf{v}_{s}[l])),\text{ }s=1,...,S.  \label{Stage_1_33}
\end{align}
\subsubsection*{\textbf{Problem 2: Optimization w.r.t.  $\bf x$ }} 
\vspace{-0.8cm}
\label{SecXmin}
\begin{align}
\small
\label{xMin}
\widehat{\bf x} =\arg \min_{\bf{x}} \frac{1}{\gamma}\sum_{s=1}^{S}||\bf{u}_{s}-\bf{A}_{s}\bf{x}||_{2}^{2} + \beta\sum_{i\in{\cal I}_p}||\bf{R}_i\bf{x}-\bf{D}{\bs \alpha }_i||_{2}^{2}.
\end{align}
The minimum condition on \eqref{xMin} obtained by equating Wirtinger derivative for complex domain function to zero, i.e., $\frac{\partial \nbr{.} }{\partial  \bf{x}}=0$, leads to the following least-square equation:
\begin{align}
\left(\sum_{s=1}^{S}\frac{\bf{A}_{s}^{H} \bf{A}_{s}}{\beta\gamma}+\sum_{i\in{\cal I}_p}\bf{R}_{i}^{H} \bf{R}_{i}\right)\bf{x}= \sum_{s=1}^{S} \frac{\bf{A}_{s}^{H}\bf{u}_{s}}{\beta\gamma}+\sum_{i\in{\cal I}_p}\bf{R}_{i}^{H} \bf{D}{\bs \alpha }_i, 
\end{align}
and to the solution:
\begin{align}
\label{bwdprop}
\widehat{\bf{x}} = & \left(\sum_{s=1}^{S}\frac{\bf{A}_{s}^{H} \bf{A}_{s}}{\beta\gamma}+\sum_{i\in{\cal I}_p}\bf{R}_{i}^{H} \bf{R}_{i}\right)^{-1} \times \left(\sum_{s=1}^{S} \frac{\bf{A}_{s}^{H}\bf{u}_{s}}{\beta\gamma}+\sum_{i\in{\cal I}_p}\bf{R}_{i}^{H} \bf{D}{\bs \alpha }_i\right).
\end{align}
We remark that \textbf{i}) the matrix  $\sum_{i\in{\cal I}_p}\bf{R}_{i}^{H} \bf{R}_{i}$ is diagonal and the $j$th diagonal element (say $\mu_j$) represents the number of times the pixel $j$ appears in  the set of extracted patches, \textbf{ii}) For the propagation matrix $\bf{A}_s=\bf{FM}_s$ (see description of \eqref{PReqn_mask}), $\bf{A}_s^H\bf{A}_s=\bf{I}_n$, where $\bf{I}_n$ is the identity matrix of order $n \times n$ and \textbf{iii}) the vector  $\sum_{i\in{\cal I}_p}\bf{R}_{i}^{H} \bf{D}{\bs \alpha }_i$ is the sum of the reconstructed patches $ \bf{D}{\bs \alpha }_i$ put back into their original locations. From the remarks \textbf{i}) and \textbf{ii}), it can be shown that the matrix to invert in \eqref{bwdprop} is diagonal with its diagonal entries given by $S/\beta\gamma+\mu_j,~~ j=1,2,...,n$. Thus the inversion of the large matrix can easily be precomputed.
\subsubsection*{\textbf{Problem 3: Optimization w.r.t.} \texorpdfstring{$\cbr{\boldsymbol{\alpha}_i}$}{\textalpha}}
\label{Secalpha}
Given that the optimization w.r.t. $\{{\bs\alpha}_i  \}$ is decoupled w.r.t. $\bs {\alpha}_i$, for $i\in{\cal I}_p$,  the optimization w.r.t.   $\{{\bs\alpha}_i \}$ amounts to compute  
\begin{align}
\label{AlphaMin}
\bs{\widehat{\alpha}}_i =\arg \min_{\bs{\alpha}} \beta||\bf{R}_i\bf{x}-\bf{D}{ \bs \alpha }||_{2}^{2}+\tau _{a} ||\bs{\alpha }||_{0}, \quad i\in{\cal I}_p.
\end{align}
Optimization   \eqref{AlphaMin} is NP-hard due to the presence of $ {l}_0$ norm \cite{1995_Natarajan_Sparse}. This issue is often circunvented by replacing the $l_0$ norm with a convex surrogate, often the $l_1$ norm, as seen in LASSO \cite{1994_Tibshirani_lasso}, BPDN \cite{1998_Chen_bpdn}, etc. However, and  in line with the findings in \cite{2015_Hongxing_Interferometric} and \cite{2009_Mairal_Online}, we use a greedy algorithm, namely the orthogonal matching pursuit (OMP) \cite{1993_Pati_OMP}, to find an approximate solution of \eqref{AlphaMin}, which is computationally light and provides  competitive results compared with those obtained by convex approximations to the $\ell_0$ norm.

The pseudo code for Complex domain OMP, courtesy of \cite{2015_Hongxing_Interferometric}, is shown in Algorithm \ref{algOMP}. 

\begin{algorithm}[h!]
	\SetKwInOut{Input}{Input}\SetKwInOut{Output}{Output}
	\caption{Orthogonal Matching Pursuit (OMP)}
	\label{algOMP}
	\Input{$\bf D \in \mathbb{C}^{w^2\times k}$ (Dictionary from C-ODL \ref{algCodl}),\\ $\bf{x}^p\in\mathbb{C}^{w^2}$, (input patch)\\ 
		$\delta > 0$ (error tolerance)}
	\Output{$\bs{\alpha}\in\mathbb{C}^k$}
	\Begin{
		$Q:=\emptyset,\alpha:=0,e:=\infty $,
		$\bs r:=\bf{x}^p$ \\
		\While{$e > \delta$}{
			$\bf u:=\bf{r}^H\bf D$\\
			$i:=\arg \max_k |\bf u_k|$ \\
			$Q:=Q \cup \cbr{i}$\\
			$\bs \beta_1 :=\bf{D}^{\#}_Q \bf{x}^p $ \\
			$\bf r :=\bf{x}^p-\bf{D}_Q \bs \beta_1 $ \\
			$e:=||\bf{r}||^2$
		}%
		$\bs \alpha\nbr{Q}:=\bs{\beta}_1$\;
	}
\end{algorithm}

In the above algorithm, $\bf{D}_Q$ is a matrix holding the atoms of $\bf{D}$ indexed by $Q$ and $\bf{D}^{\#}_Q$ is the pseudo-inverse of $\bf{D}_Q$. Also, $\bs \alpha\nbr{Q}$ represents the components of $\bs \alpha$ with indexes in $Q$.


\subsubsection*{\textbf{Problem 4: Optimization w.r.t. $\bf D$ }} 
\label{secDmin}
If the dictionary $\bf D$ is  known beforehand,  the optimization w.r.t. $\bf D$ is disregarded. However, if $\bf D$ is unknown beforehand, we should solve the quadratically constrained quadratic program (QCQP)   
\begin{align}
\label{eq:D_opt}
\min_{\bf{D}} & \sum_{i\in{\cal I}_p}||\bf{R}_i\bf{x}-\bf{D}{\bs \alpha }_i||_{2}^{2}\\ 
\text{s.t.:}& ~~{\bf D}\in \mathcal{C}:=\cbr{\bf{D} \in \mathbb{C}^{w^2 \times k}: \|{\bf d}_j\|^2\leq 1,\, \forall j=1,\dots, k},
\end{align}
where the convex set  $\mathcal{C}$ is introduced to prevent the dictionary atoms $\cbr{\bf{d}_j}_{j=1}^k$ being arbitrarily large, which in turn leads to arbitrarily small values of the codes ${\bs \alpha}_i$. 
We remark that the presence of the $\ell_0$ terms $\sum_{i\in {\cal I}_p} \|{\bs \alpha}_i\|_0$ in  the objective function $\cal L$ promotes sparse codes and, therefore, as desired,  dictionaries are able to sparsely represent the restored patches. In a large number of experiments, we have observed, however, that the quality of the dictionary, regarding its ability to produce sparse codes, improves if we include an 
$\ell_1$ norm in the optimization \eqref{eq:D_opt} and solve simultaneously w.r.t. $\bf D$ and $\bs \alpha$, that is, if we solve,  
\begin{align}
\label{eq:D_opt_l1}
\min_{\bf{D}, {\bs \alpha}} & \sum_{i\in{\cal I}_p}||\bf{R}_i\bf{x}-\bf{D}{\bs \alpha }||_{2}^{2}+\lambda ||\bs{\alpha }||_{1}\\ \nonumber
\text{s.t.:}& ~~{\bf D}\in \mathcal{C}.
\end{align}
These findings are  in line with those observed in \cite{2009_Mairal_Online} and \cite{2015_Hongxing_Interferometric}. 

In \eqref{eq:D_opt_l1}, the joint  optimization  with respect to $\bf{D}$ and $\bs \alpha$ is not  convex, although it is  convex with respect to each variable alone. This naturally invokes the alternating minimization methodology, in which one variable is minimized at a time keeping the other one fixed, in an iterative manner. 

The real domain version of the optimization problem \eqref{eq:D_opt_l1} was solved in  \cite{2009_Mairal_Online} using the alternating minimization framework. More recently, its complex domain version was proposed for interferometric phase image estimation \cite{2015_Hongxing_Interferometric} with slight modifications, which we term as C-ODL. Both ODL and C-ODL uses BPDN for sparse coding, i.e., the optimization w.r.t. $\bs{\alpha}$, and projected block-coordinate descent method to update the columns of the dictionary. The BPDN problem in \cite{2009_Mairal_Online} is solved using Least Angle Regression (LARS) \cite{2004_Efron_LARS}, whereas in \cite{2015_Hongxing_Interferometric} a much faster algorithm called sparse regression by variable splitting and augmented Lagrangian (SpaRSAL) is introduced for that purpose. We adopt the dictionary learning methodology proposed in C-ODL, whose pseudo code, courtesy of \cite{2015_Hongxing_Interferometric}, is shown in Algorithm \ref{algCodl}. 

\begin{algorithm}[h!]
	\SetKwInOut{Input}{Input}\SetKwInOut{Output}{Output}\SetKwRepeat{Do}{do}{while}
	\caption{\footnotesize Complex Domain Online Dictionary Learning (C-ODL) }
	\label{algCodl}
	\Input{$\bf x_i^p \in \mathbb{C}^{w^2}, i\in {\cal I}_p$ (patches for dictionary learning),\\
		$T\in\mathbb{N}$ (number of iterations),\\
		$\eta \in\mathbb{N}$ (number of patches per iteration),\\ 
		$\lambda > 0 $ (BPDN regularization parameter),\\
		$\beta_t$ (damping sequence \footnotemark ),\\
		$\bf{D}_0 \in \mathbb{C}^{w^2\times k}$ (initial dictionary).}
	\Output{$\bf{D}\in\mathbb{C}^{w^2\times k}$ (estimated dictionary).}
	\Begin
	{
		$\bf{D}:=[\bf{d}_1,...,\bf{d}_k]=\bf{D}_0$\\
		$\bf{A}:=[\bf{a}_1,...,\bf{a}_k]=\bf 0$, $\bf{B}:=[\bf{b}_1,...,\bf{b}_k]=\bf 0$\\
		\For {$t \leq T$} {
			Select randomly $\bf{x}^{p,t} \equiv \sbr{\bf{x}^{p,t}_{\gamma_j}, j=1,...,\eta}$, $\gamma_j$'s are randomly picked up from ${\cal I}_p$\ \hspace{3.5 cm}
			/* Optimize wrt $\bs{\alpha}$: BPDN */ \\
			$\bs{\alpha}^t=\underset{\bs{\alpha \in \mathcal{C}^{k\times \eta}}}{\arg \min}  \frac{1}{2} ||\bf{x}^{p,t}-\bf{D}{\bs \alpha }||_{2}^{2}+\lambda||\bs{\alpha }||_{1}$\\
			$\bf{A}:=\beta_t\bf{A}+\sum_{i=1}^{\eta}{\bs \alpha}_i^t\nbr{{\bs \alpha}_i^t}^H$\\
			$\bf{B}:=\beta_t\bf{B}+\sum_{i=1}^{\eta}{\bf x}_i^{p,t}\nbr{{\bs \alpha}_i^t}^H$\ \hspace{9.0 cm}
			/* Optimize wrt $\bf{D}$: BCD */ \\
			\Repeat{convergence}{
				\For{$l=1$ {to} $l=k$}{
					$\bf u_l:=\frac{1}{\bf{A}\nbr{l,l}}\nbr{\bf b_l-\bf{D} \bf a_l}+\bf d_l$ \\
					$\bf  d_l:=\frac{\bf u_l}{\text{max} \cbr{||\bf  u_l||_2,1}}$\\
				}
			}
		}    
	}
\end{algorithm}

\subsection{DLPR for the Gaussian Observation model} 
\label{sec:Gausdervn}
We now derive DLPR for the Gaussian noise as detailed in \eqref{GaussianNoise}. Then, the likelihood function in \eqref{DLPR} can be rewritten as 
\begin{align}
\label{Gliklihood}
\mathcal{L}(\{\bf{u}_{s}\}\text{, }\bf{x}\text{,}\{\bs{\alpha}_i\}\text{, }\bf{D})=\frac{1}{\sigma ^{2}}\sum_{s=1}^{S}\sum_{l=1}^{n}[|\mathbf{u}_{s}[l]|^{2}-\mathbf{z}_{s}[l]]^{2} +\frac{1}{\gamma }||\bf{u}_{s}-\bf{A}_{s}\bf{x}||_{2}^{2}+\sum_{i\in{\cal I}_p}\tau_{a}||\bs{\alpha}_i||_{0}+\beta||\bf{R}_i\bf{x}-\bf{D}{\bs \alpha }_i||_{2}^{2}.
\end{align}
Comparing \eqref{DLPR} and \eqref{Gliklihood}, it is clear that DLPR optimization formulations are the same w.r.t. the variables   $\bf x$, $\{\bs \alpha_i \}$, and $\bf D$,  which are derived in the Sections \ref{SecXmin}, \ref{Secalpha} and \ref{secDmin} respectively. The only difference is in  the amplitude update at the sensor plane, i.e., the optimization w.r.t  $\{\bf{u}_{s}\}$. Minimization w.r.t. $\bf{u}_s$ on \eqref{Gliklihood} leads to the following Cardan equation, whose solution yields $\mathbf{\widehat{b}}_{s}[l]$ \cite{2016_Soulez_Proximity}, \cite{2012_Katkovnik_mask}, \cite{2017_Katkovnik_Phase_Retrieval}: 
\begin{align}
&\mathbf{b}_{s}^{3}[l]+C\mathbf{b}_{s}[l]+D=0\text{,}
\label{bslGau} \nonumber \\
& C=\frac{\sigma ^{2}}{2\gamma }-{\bf z}_{s}[l]\text{, }D=-\frac{\sigma ^{2}}{2\gamma}|\mathbf{v}_{s}[l]|. 
\end{align}
We remark here that, as the value of $\gamma$ changes from $0$ to $\infty$, the solution of \eqref{bslGau} changes from $|\mathbf{v}_{s}[l]|$ to $\sqrt{\bf{z}_{s}[l]}$.

\begin{figure}[h!]
	\centering
	\includegraphics[width=1.0\textwidth]{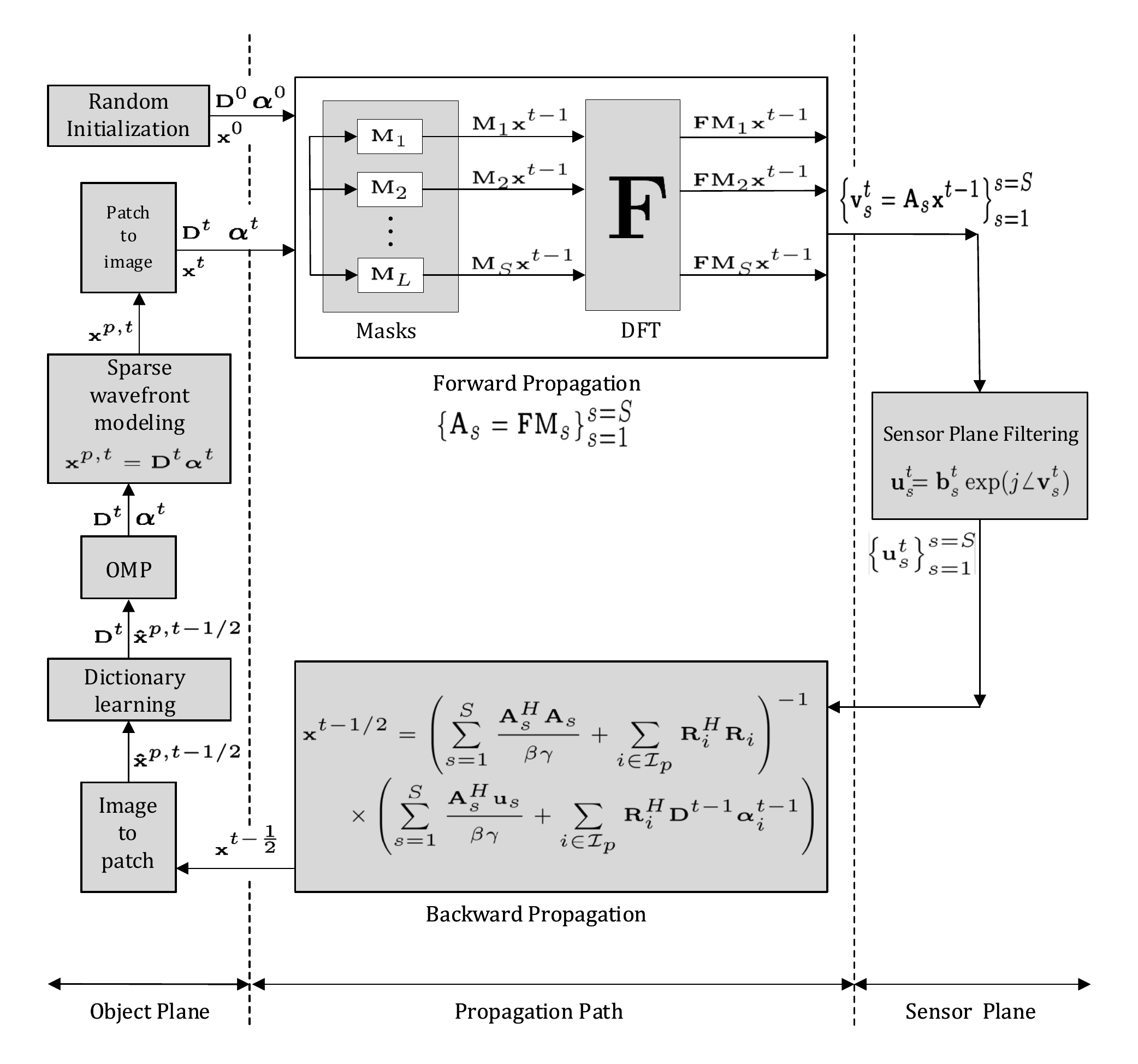}
	\caption{Block Diagram of the DLPR algorithm}
	\label{dlpr_bd}
\end{figure}

Pseudo code for the proposed  DLPR algorithm, which combines the results derived in \ref{sec:Poisdervn} and \ref{sec:Gausdervn} for forward propagation, sensor plane filtering, backward propagation, and object plane sparsification  is given in Algorithm \ref{DLPRalg}. A schematic representation of the DLPR algorithm in the form of a block diagram is shown in Fig. \ref{dlpr_bd}. To better understand the signal flow of DLPR in relation with the optical setup \ref{setup_mask}, the block diagram is partitioned into 3 regions: \textbf{i}) Object plane - All the operations related to the wavefront sparse modeling, namely patch formation, dictionary learning, OMP, sparse regression and patch aggregation are part of this region. These blocks perform the operation given by the steps- 5, 6, 7, 8 \& 9 of Algorithm \ref{DLPRalg}. \textbf{ii}) Propagation path- This is the middle region of the schematic representation and models the forward and backward optical propagations described by step-2 and step-4 of Algorithm \ref{DLPRalg}. \textbf{ii}) Sensor plane- The filtering operations at the sensor plane (step-3 of Algorithm \ref{DLPRalg}) for Poissonian  and Gaussian observations are included in this part.
\footnotetext{For the damping sequence $\beta_t$, we use the implementation from \cite{2009_Mairal_Online}. A detailed description of this parameter is beyond the scope of our  discussion.}
\begin{algorithm}[h!]
	\SetKwInOut{Input}{Input}\SetKwInOut{Output}{Output}\SetKwInOut{Init}{Init}
	\caption{Dictionary Learning Phase Retrieval (DLPR)}
	\label{DLPRalg}
	\Input{$\mathbf{z}_{s}\in \mathbb{R}^{n}$ (noisy observation)\\
		\Init{$\widehat{\mathbf{x}}^0 \in \mathbb{C}^{n}$ (object wavefront), \\ $\mathbf{D}^{0}\in \mathbb{C}^{w^2\times k}$ (dictionary), \\ $\bs{\alpha}^{0}\in \mathbb{C}^{k}$ (code), \\$\gamma,\beta>0$(regularization  parameter), \\$S\in\mathbb{N}$ (Number of observations),  \\ $T\in\mathbb{N}$ (number of iteration)}}
	\Output{$\widehat{\mathbf{x}}^T, \bs{\alpha}^T, \bf{D}^T$}
	\For{$t\leq T$}{
		\textbf{Forward propagation:} \hspace{10cm}
		$\mathbf{\hat{v}}_{s}^{t}=\mathbf{A}_{s}\mathbf{\hat{x}}^{t-1}\text{, where } \mathbf{A}_{s}=\mathbf{FM}_{s}$ and $s=1,...,S$\\
		\textbf{Filtering at Sensor Plane:}\hspace{10cm}
		 $\mathbf{\hat{u}}_{s}^{t}=\mathbf{\hat{b}}_{s}^{t}\mathbf{\odot }\exp \nbr{j\cdot \text{angle}\nbr{\mathbf{\hat{v}}_{s}^{t}}}, s=1,...,S, $ where $\mathbf{\hat{b}}_{s}^{t}$ is given by \eqref{bslPoi} and \eqref{bslGau} for Poissonian and Gaussian observations respectively\\
		\textbf{Backward propagation:} 
		$\bf{\hat{x}}^{t-1/2}=  \left(\sum_{s=1}^{S}\frac{\bf{A}_{s}^{H} \bf{A}_{s}}{\beta\gamma}+\sum_{i\in{\cal I}_p}\bf{R}_{i}^{H} \bf{R}_{i}\right)^{-1} \times  \left(\sum_{s=1}^{S} \frac{\bf{A}_{s}^{H}\bf{u}_{s}}{\beta\gamma}+\sum_{i\in{\cal I}_p}\bf{R}_{i}^{H} \bf{D}^{t-1}{\bs \alpha }_i^{t-1}\right)$, from \eqref{bwdprop}\\
		\textbf{Patch formation:}  \hspace{10cm}
		$\bf{\hat{x}}_i^{p, t-1/2}= {\bf R}_i\bf{\hat{x}}^{t-1/2},~\forall i\in {\cal I}_p$ \\
		\textbf{Dictionary learning:}  \hspace{10cm}
		Obtain $\bf{D}^t$ by running C-ODL (Algorithm \ref{algCodl}) iterations using $\bf{\hat{x}}_i^{p, t-1/2}$ as input\\
		\textbf{Sparse code learning:}  \hspace{10cm}
		Obtain $\bs{\alpha}_i^t,~\forall i\in {\cal I}_p$ by running OMP (Algorithm \ref{algOMP}) iterations using $\bf{\hat{x}}_i^{p, t-1/2}$ and $\bf{D}^t$ as input\\
		\textbf{Sparse wavefront modeling at the object plane:}  \hspace{10cm}
		$\widehat{\bf{x}}_i^{p,t}=\bf{D}^t{\bs {\alpha}}_i^t,~\forall i\in {\cal I}_p$\\
		\textbf{Patch to image formation:}  \hspace{4cm}
		$\widehat{\bf{x}}^t=\nbr{\sum_{i\in{\cal I}_p}\bf{R}_{i}^{H} \bf{R}_{i}}^{-1}\nbr{\sum_{i\in{\cal I}_p}\bf{R}_{i}^{H}\widehat{\bf{x}}_i^{p,t}} $ from\eqref{p2imag}
	}
\end{algorithm}

\section{Experiments and Results}
\label{SecExp}
In this section, we present a series of strong empirical evidences, using real and synthetic data, to illustrate the effectiveness of DLPR. Prior to the results, the details regarding the experimental setup are described below:

\textbf{Optical setup:}
We restrict our optical setup to the lenseless imaging scenario, as illustrated in Fig. \ref{setup_mask}, which uses coded diffraction patterns for optical imaging. To implement the wavefront modulation, in alignment with the works presented in \cite{2015_Candes_PRWF}, \cite{2015_Chen_TWF} and \cite{2017_Katkovnik_Phase_Retrieval}, a random phase value sampled from the set $\cbr{0, \pi /2, -\pi /2, \pi}$ is selected with equal probability for each pixel of the mask.

\textbf{State-of-the-art competitors:}
The proposed DLPR algorithm is designed mainly for moderate and highly noisy scenarios. The recent TWF algorithm \cite{2015_Chen_TWF}, designed for noiseless or small level noisy data, is chosen for comparison purposes. To the best of our knowledge, SPAR \cite{2017_Katkovnik_Phase_Retrieval} is the state-of-the-art for retrieving phase from the noisy observations and is a good candidate for the comparisons. We also consider a third candidate, which we term as GS-F algorithm, by skipping the sparse modeling of the wavefront at the object plane, i.e., by making the substitution $\bf{\hat{x}}^{t}=\bf{\hat{x}}^{t-1/2}$, in Algorithm \ref{DLPRalg}.The main rationale behind GS-F is that it helps to visualize how much improvement, in terms of performance, is contributed by the proposed sparse wavefront modeling. The performance of the conventional GS algorithm is expected to lag behind GS-F in noisy observation scenarios, as the GS lacks sensor plane filtering.

Although we consider SPAR, TWF, and GS-F for comparison purposes, the main competitor of DLPR is SPAR as it includes sparsity based object wavefront filtering aiming at good performance for noisy data. Since the SPAR algorithm applies sparsity on absolute phase, phase unwrapping has to be implemented to obtain the absolute phase in each iteration, which sets a bottleneck to the performance of SPAR, especially for highly noisy data. We make use of the publicly available MATLAB demo-codes\footnote{http://www.cs.tut.fi/sgn/imaging/sparse/} for SPAR and TWF.

\textbf{DLPR parameter settings:}
Parameter settings are crucial in DLPR's performance. The parameters of Algorithm \ref{DLPRalg} are heuristically set to the following values: $\gamma=1/\chi,\beta=\chi/1000$ for Poissonian observations and $\gamma=\sigma^2/10,\beta=0.01/\sigma^2$ for Gaussian observations. For the wavefront modulation, SPAR \cite{2017_Katkovnik_Phase_Retrieval} suggests a value of $S\geq10$. Without loss of generality, in alignment with the  suggestions provided in \cite{2017_Katkovnik_Phase_Retrieval}, all our experiments are conducted by keeping $S = 12$ , i.e., for 12 observations with different phase modulation masks. The dictionary learning (C-ODL) \ref{algCodl} and OMP \ref{algOMP} sub-iterations are tuned to their best performance as per the parameter settings described in \cite{2015_Hongxing_Interferometric}. The patches used in all experiments are square, having dimension $10\times10$, extracted with unit stride. 

We also remark that the algorithms chosen for comparison, i.e., SPAR, GS-F and TWF, are tuned to their optimal performances by using the parameter settings given in \cite{2017_Katkovnik_Phase_Retrieval}.

\textbf{Performance evaluation:}
Since the phase is the main focus of a typical PR problem, the performance of DLPR is evaluated based on the quality of the retrieved phase $\widehat{\bs{\psi}}_{2\pi}$ using  Root Mean Square Error ($\text{RMSE}$) defined as
\begin{eqnarray}
\text{RMSE}_{\bs{\psi} }:=\sqrt{||\mathcal{W}\nbr{\widehat{\bs{\psi}}_{2\pi}-\bs{\psi}
		_{2\pi}}||_{2}^{2}},
\end{eqnarray}
where $\mathcal{W}$ is the wrapping operator defined is \eqref{eqn:wrap} and $\widehat{\bs{\psi}}_{2\pi}$ is the estimate of the true wrapped phase  $\bs{\psi}_{2\pi}$.

\begin{figure}[h!]
	\centering
		\begin{subfigure}{0.19\textwidth}
		\includegraphics[width=\textwidth]{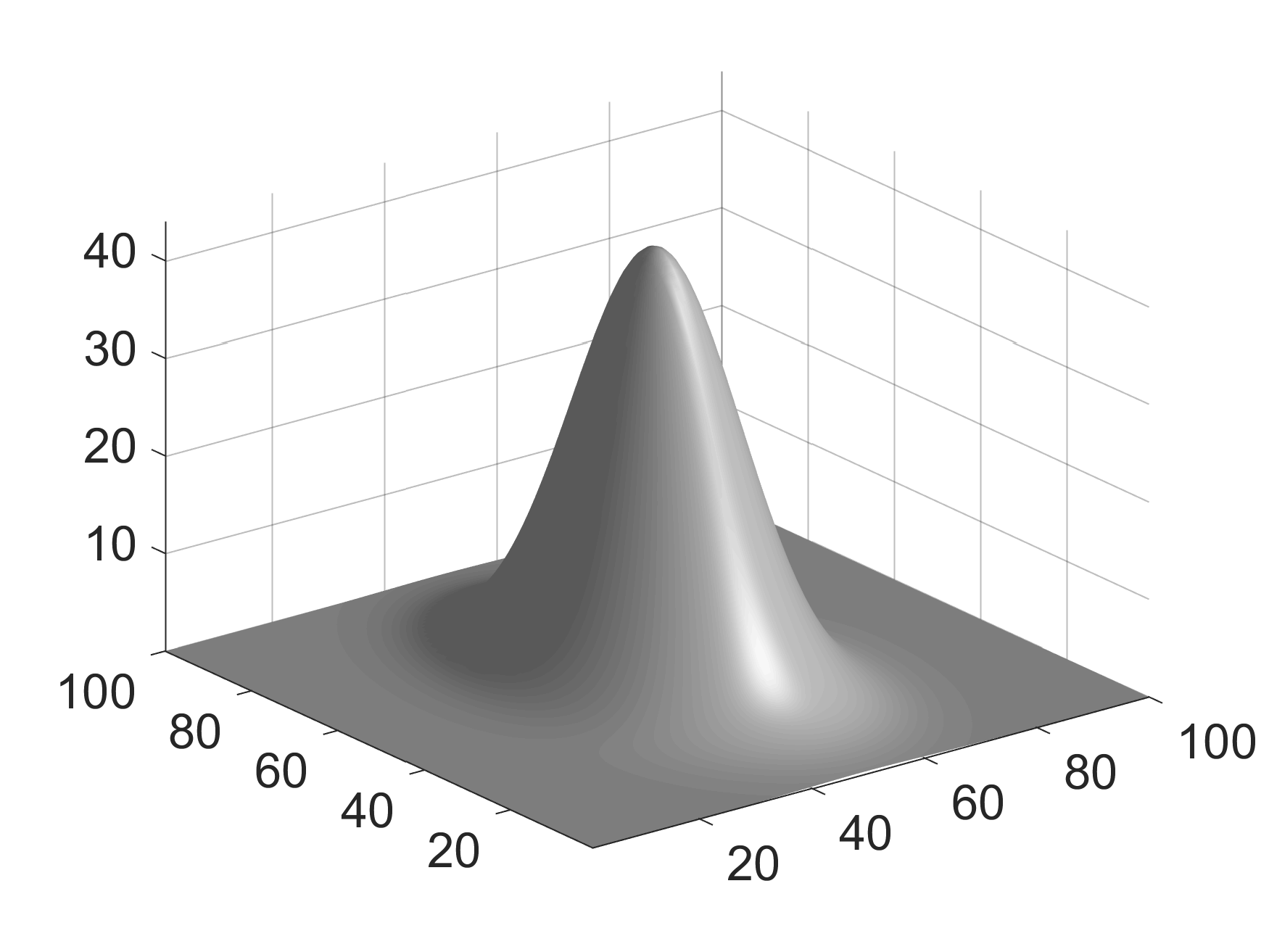}
		\caption{Gaussian}
		\label{TG}
	\end{subfigure}
	\hfill
	\begin{subfigure}{0.19\textwidth}
		\includegraphics[width=\textwidth]{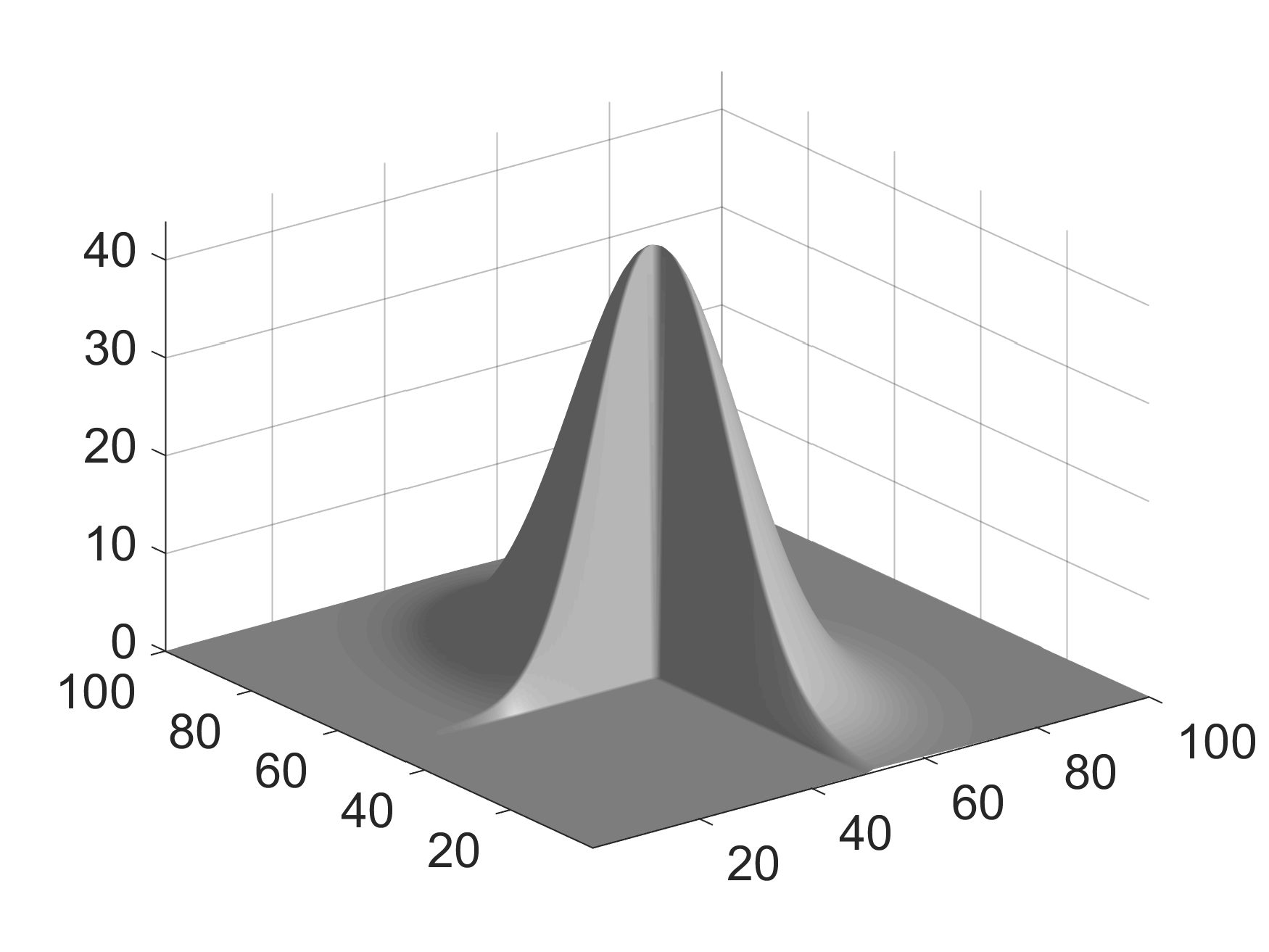}
		\caption{Trunc. Gaussian}
		\label{TG}
	\end{subfigure}
	\hfill
	\begin{subfigure}{0.19\textwidth}
		\includegraphics[width=\textwidth]{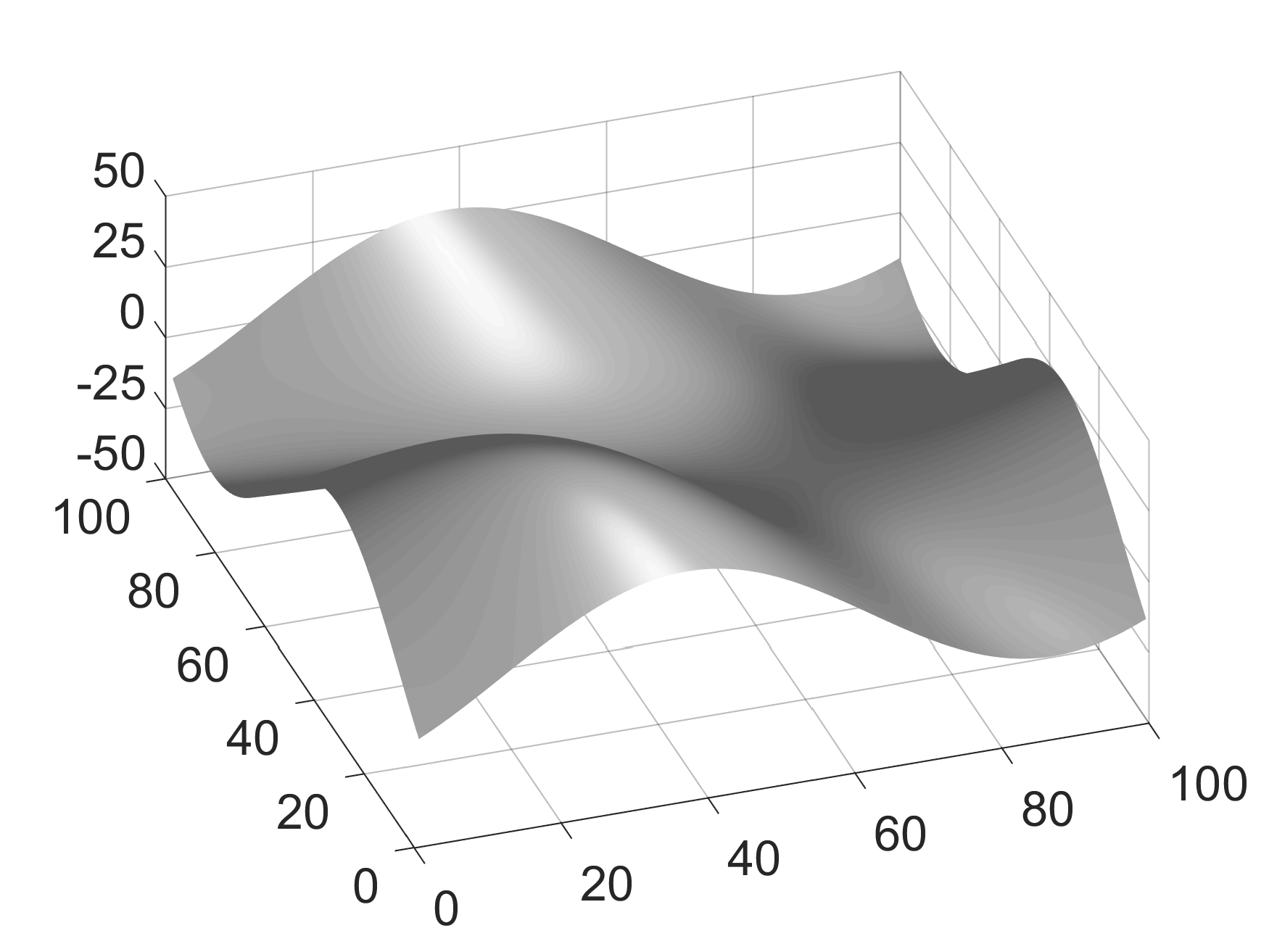}
		\caption{Mountain}
		\label{MNT}
	\end{subfigure}
	\hfill
	\begin{subfigure}{0.19\textwidth}
		\includegraphics[width=\textwidth]{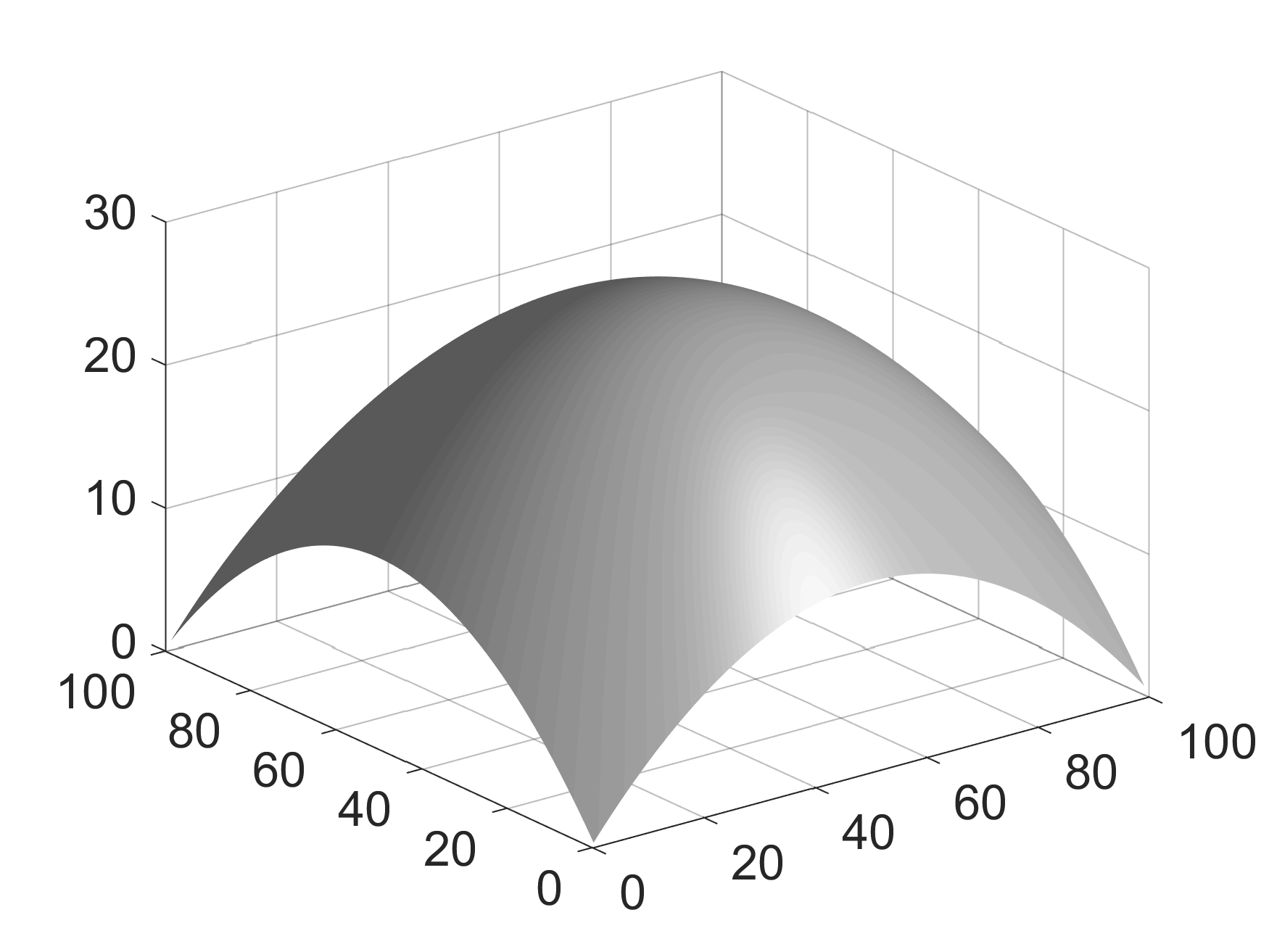}
		\caption{Quadratic Surface}
		\label{quad}
	\end{subfigure}
	\hfill
	\begin{subfigure}{0.19\textwidth}
		\includegraphics[width=\textwidth]{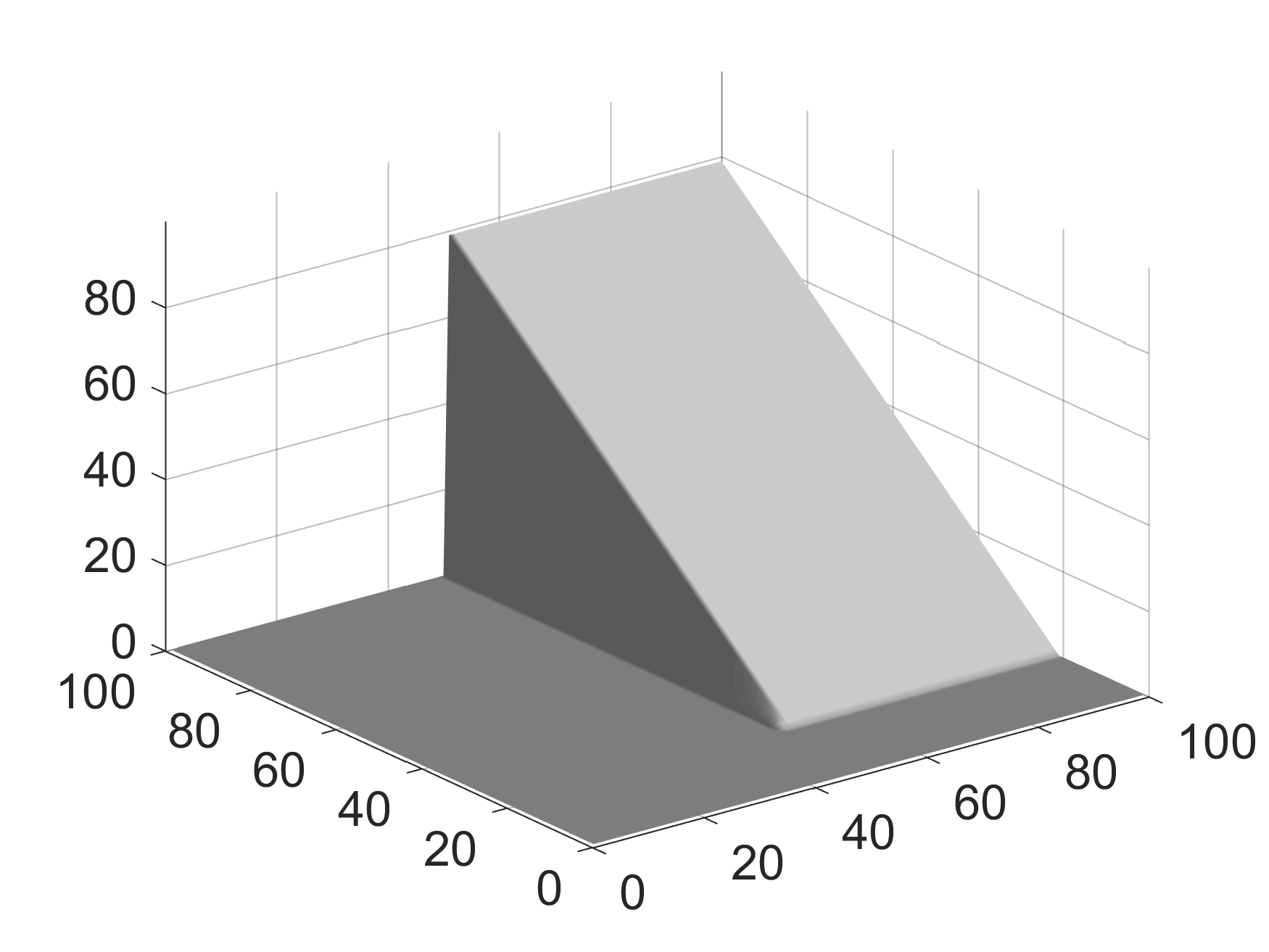}
		\caption{Shear Plane}
		\label{SP2}
	\end{subfigure}
	\caption{Synthetic phase surfaces to build the complex valued signal.}
	\label{SimData}
\end{figure}

\textbf{Synthetic data set:}
In an optical wavefront $\bf{x=a\odot}e^{j\bs{\psi}}\in\mathbb{C}^n$, the vectorised phase image $ \bs{\psi}\in\mathbb{R}^n$ and amplitude image $\bf{a}\in\mathbb{R}_+^n$ can be quite correlated. In order to simulate various amplitude-phase relationships, we create different groups of complex signals in which the phases and amplitudes are related in different ways. We introduce the following formula to relate phase and amplitude in different ways:
\begin{align}
\bf{a}_i:=k_0+k_1f\nbr{\bs{\psi}_i},\label{sig_group}
\end{align}
where $k_0$ and $k_1$ are parameters used to control the level of amplitude and the function $f\nbr{.}$ is used to control the correlation between phase and amplitude. Based on \eqref{sig_group}, we define the following groups of complex signals:
\begin{enumerate}
	\item Group-1: Invariant amplitude, i.e., $\bs{a}_i=1$.
	\item Group-2: Independent amplitude, i.e., amplitude and phase are two unrelated images. 
	\item Group-3: Amplitude and phase are highly similar. $ k_0=1, k_1=1, f\nbr{\bs{\psi}_i}:=\frac{|\bs{\psi}_i|}{\text{max} (|\bs{\psi}_i|)}$.
	\item Group-4: Amplitude and phase are less similar. $ k_0=1, k_1=1, f\nbr{\bs{\psi}_i}:=|{\rm cos}(15\bs{\psi}_i)|$.
\end{enumerate}  
Using the above definitions for the image amplitudes and the simulated phase surfaces as shown in Fig. \ref{SimData}, nine different complex valued synthetic data are generated, which are summarised in Table \ref{tab:testsig}.

\begin{table}[h!]
	\centering
	\caption{Simulated complex images}
	\label{tab:testsig}
	\begin{tabular}{c|l|l|c}
		Sig No & amplitude $\bf a$ & phase $\boldsymbol {\psi}$ & \multicolumn{1}{l}{Group} \\
		\midrule
		\midrule
		1     & constant & Trun. Gaussian & \multirow{2}[2]{*}{1} \\
		2     & constant & Shear Plane &  \\
		\midrule
        \midrule
		3     & Mountain & Shear Plane & \multirow{3}[2]{*}{2} \\
		4     & Quadratic & Trun. Gaussian &  \\
		5     & Gaussian & Shear Plane &  \\
		\midrule
        \midrule
		6    & Highly similar & Trun. Gaussian & \multirow{2}[2]{*}{3} \\
		7    & Highly similar& Shear Plane &  \\
		\midrule
        \midrule
		8    & Less similar& Trun. Gaussian & \multirow{2}[2]{*}{4} \\
		9    & Less similar & Shear Plane &  \\
		\bottomrule
		\bottomrule
	\end{tabular}
\end{table}
\subsection{Poissonian Observations}
\subsubsection{Experiments using synthetic data set}
\begin{figure}[h!]
	\centering
	\begin{subfigure}{0.49\textwidth}
		\includegraphics[width=\textwidth]{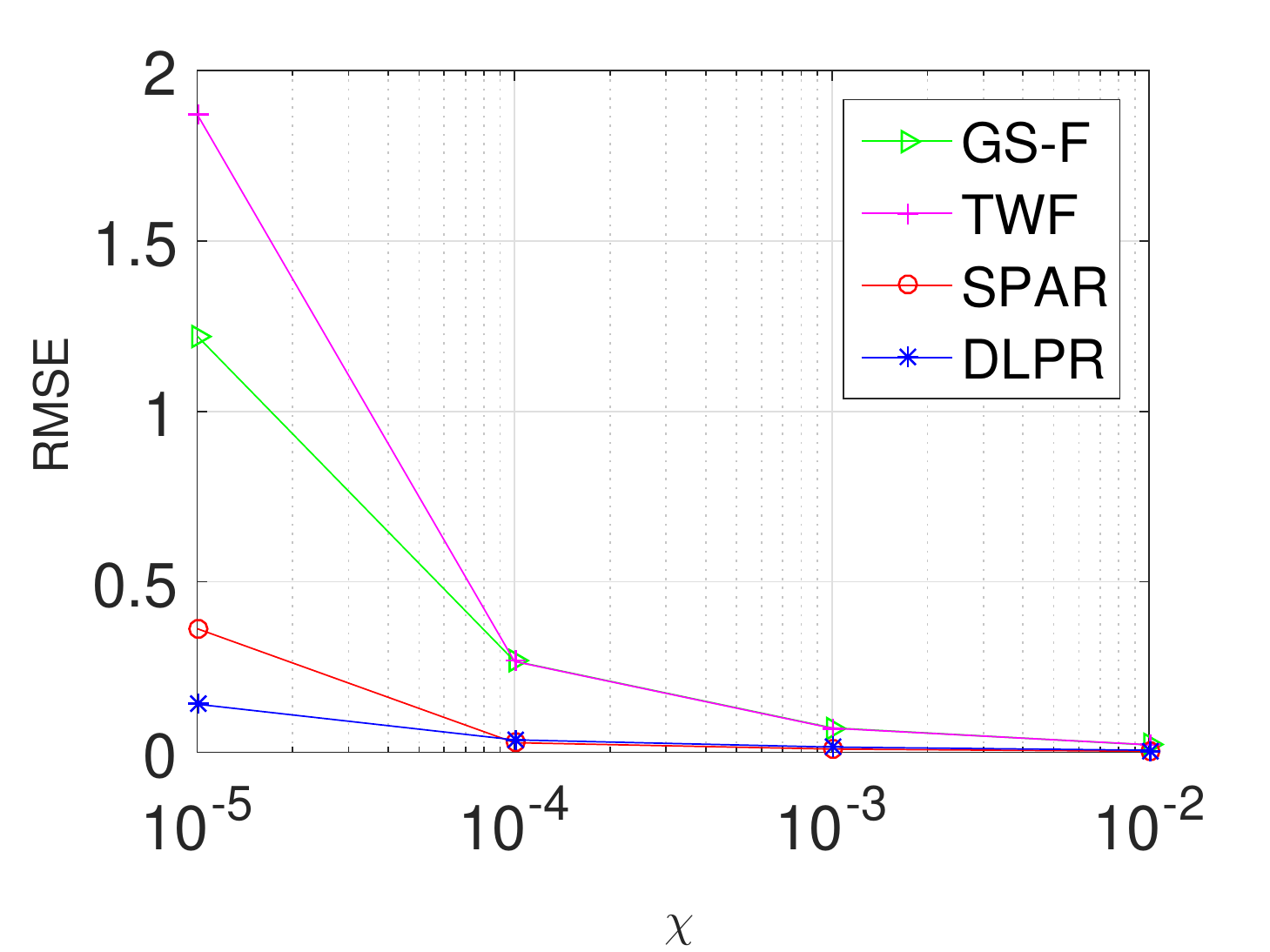}
		\caption{Group-1}
		\label{pgp1}
	\end{subfigure}
\vspace{0.1cm}
	\hfill
	\begin{subfigure}{0.49\textwidth}
		\includegraphics[width=\textwidth]{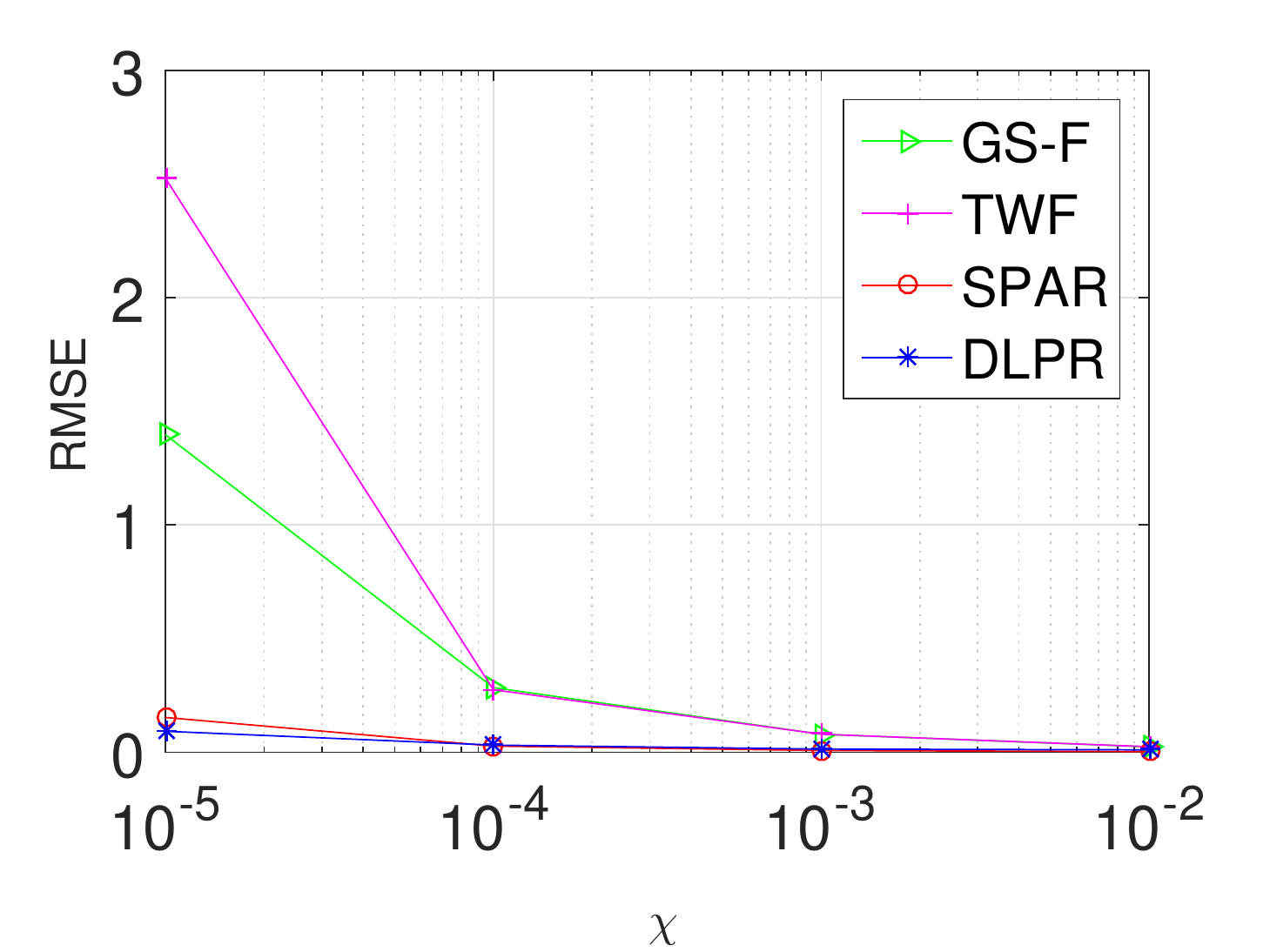}
		\caption{Group-2}
		\label{pgp2}
	\end{subfigure}
\vspace{0.1cm}
	\hfill
	\begin{subfigure}{0.49\textwidth}
		\includegraphics[width=\textwidth]{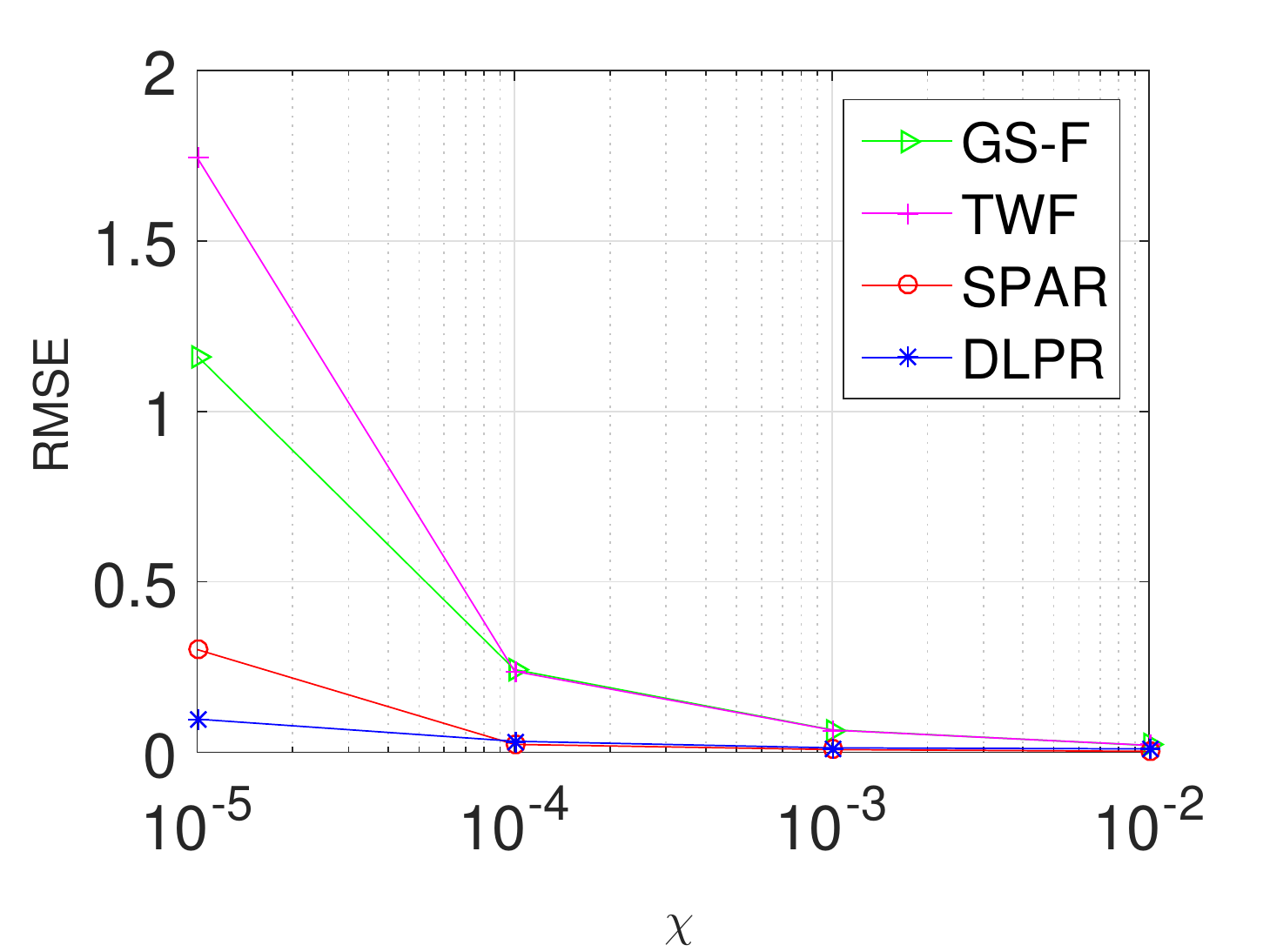}
		\caption{Group-3}
		\label{pgp3}
	\end{subfigure}
	\hfill
	\begin{subfigure}{0.49\textwidth}
		\includegraphics[width=\textwidth]{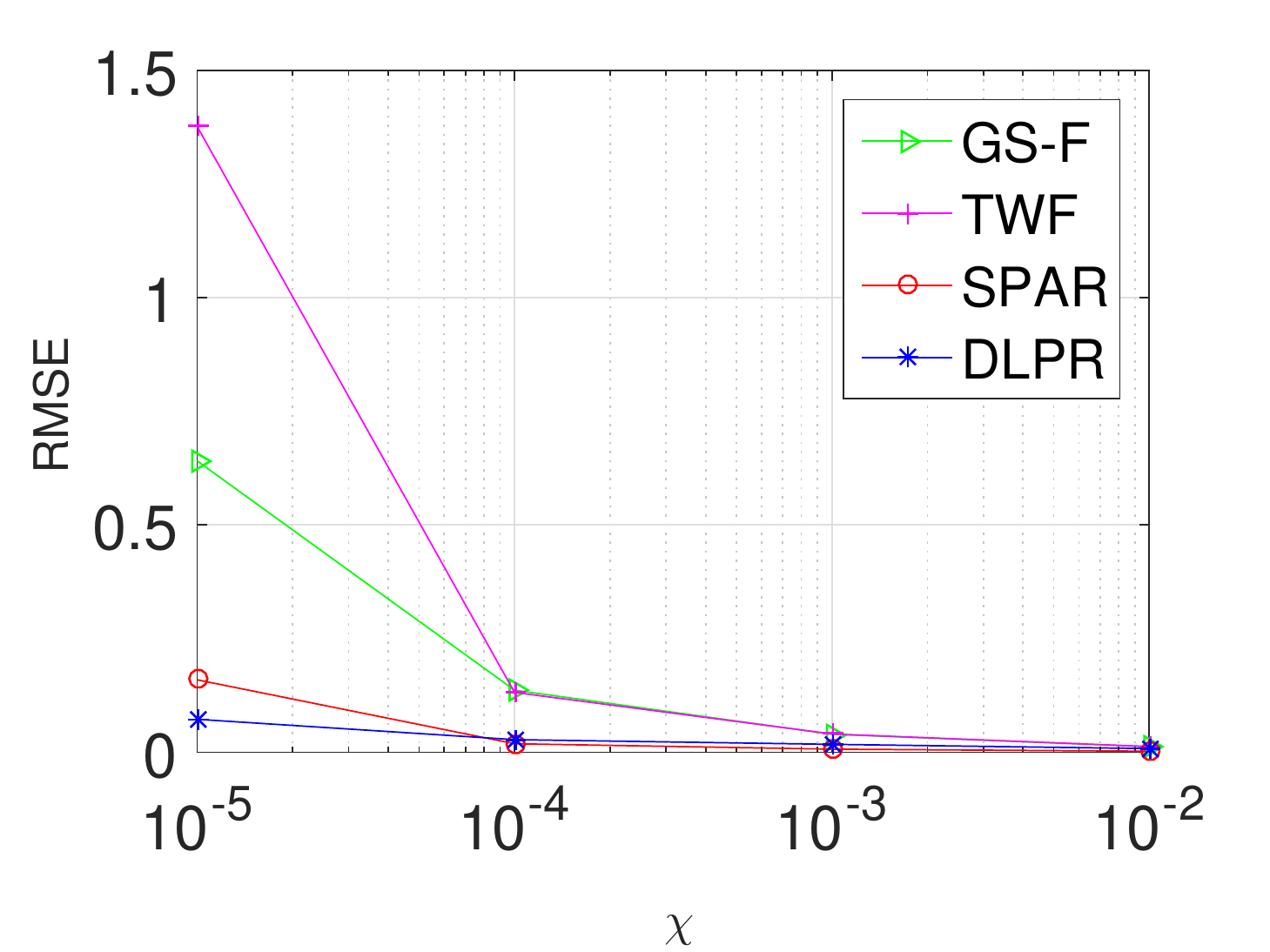}
		\caption{Group-4}
		\label{pgp4}
	\end{subfigure}
	\caption{RMSE of retrieved phase using DLPR in comparison with SPAR, TWF, GSF. Poissonian observations model is considered.}
	\label{RMSE_Pois}
\end{figure}

In this section, the experiments conducted using the data set given in Table \ref{tab:testsig} are presented. DLPR is tested for each group in the Table \ref{tab:testsig} by considering noise power ranging from high to medium level, i.e., $\chi = \sbr{0.00001,~0.0001,~0.001,~0.01} $. SNR values corresponding to these noise levels are respectively $\sbr{-7,~3,~13,~23}$ dB. RMSE values of the retrieved phase, averaged for each group, are given in Fig. \ref{RMSE_Pois}. The performances of SPAR, TWF and GS-F are also included in the same figure. It is evident from the figure that the DLPR shows remarkable improvement over TWF and GS-F. In comparison with SPAR, DLPR performs better for heavy noisy data ($\chi = 0.00001$). This observation is in clear alignment with our early intuition about DLPR. On the other hand, for medium level of noise, DLPR and SPAR give very close results. It should be noted that the low-noise scenario is not considered here as the objective of our algorithm is to retrieve phase from highly noisy observations. However, we remark that when the observation mechanism is noise free or with very low level of noise, all the four algorithms show similar performances.
\subsubsection{Phase unwrapping}
\begin{figure}[h!]
	\centering
	\begin{subfigure}{0.49\textwidth}
		\includegraphics[width=\textwidth]{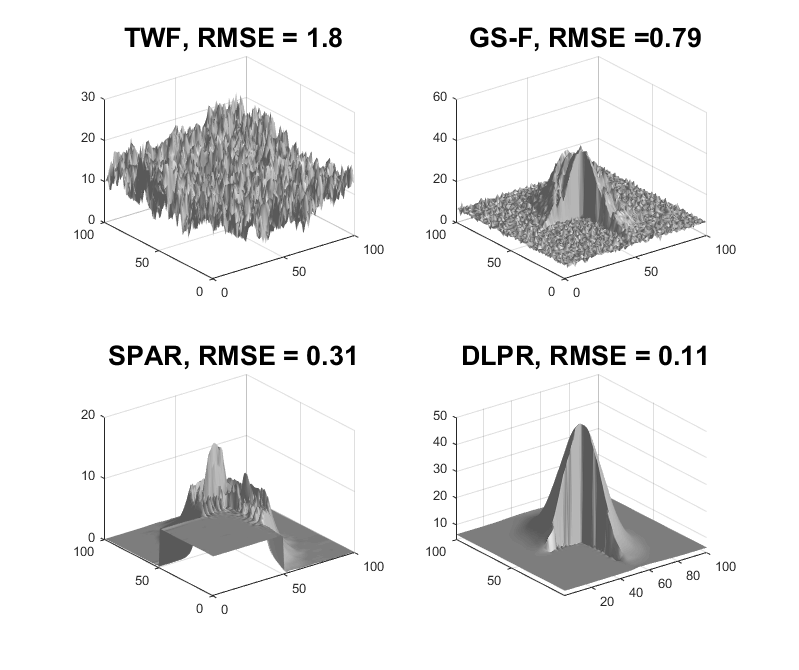}
		\caption{  Truncated Gaussian }
		\label{puTG}
	\end{subfigure}
	\hfill
	\begin{subfigure}{0.49\textwidth}
		\includegraphics[width=1.1\textwidth]{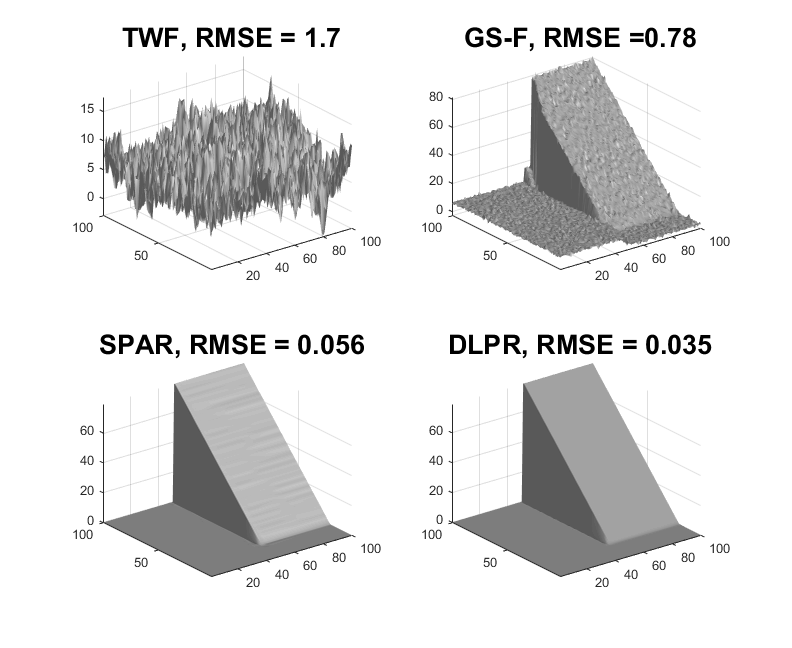}
		\caption{  Shear plane}
		\label{puSP}
	\end{subfigure}
	\caption{ Absolute phase estimation. Poissonian observations with high noise level ($\chi=0.0001/4,~{\rm SNR}=-0.1~{\rm dB}$) is considered.}
\end{figure}

In a practical PR scenario, the phase obtained from retrieved complex wavefront is wrapped. The absolute phase is obtained by \textit{phase unwrapping} in which the quality of the retrieved wrapped phase is a crucial factor. The following experiment is a qualitative illustration to show that the high quality of the phase images retrieved through DLPR underlies the good results during the unwrapping step. Here the phase retrieved from heavily noisy observations are unwrapped using PUMA algorithm \cite{2007_Bioucas_Phase} (state-of-the-art in phase unwrapping). It is evident from Figs. \ref{puTG} \& \ref{puSP} that TWF and GS-F provide very poor estimates. Compared to these two algorithms, DLPR and SPAR perform better. For shear plane \ref{puSP}, DLPR estimate is slightly better compared to SPAR. But for truncated Gaussian \ref{puTG}, DLPR is much better than SPAR.
\subsubsection{Experiments using real MRI interferograms}
\begin{figure}[h!]
	\centering
	\begin{subfigure}{0.32\textwidth}
		\includegraphics[width=\textwidth]{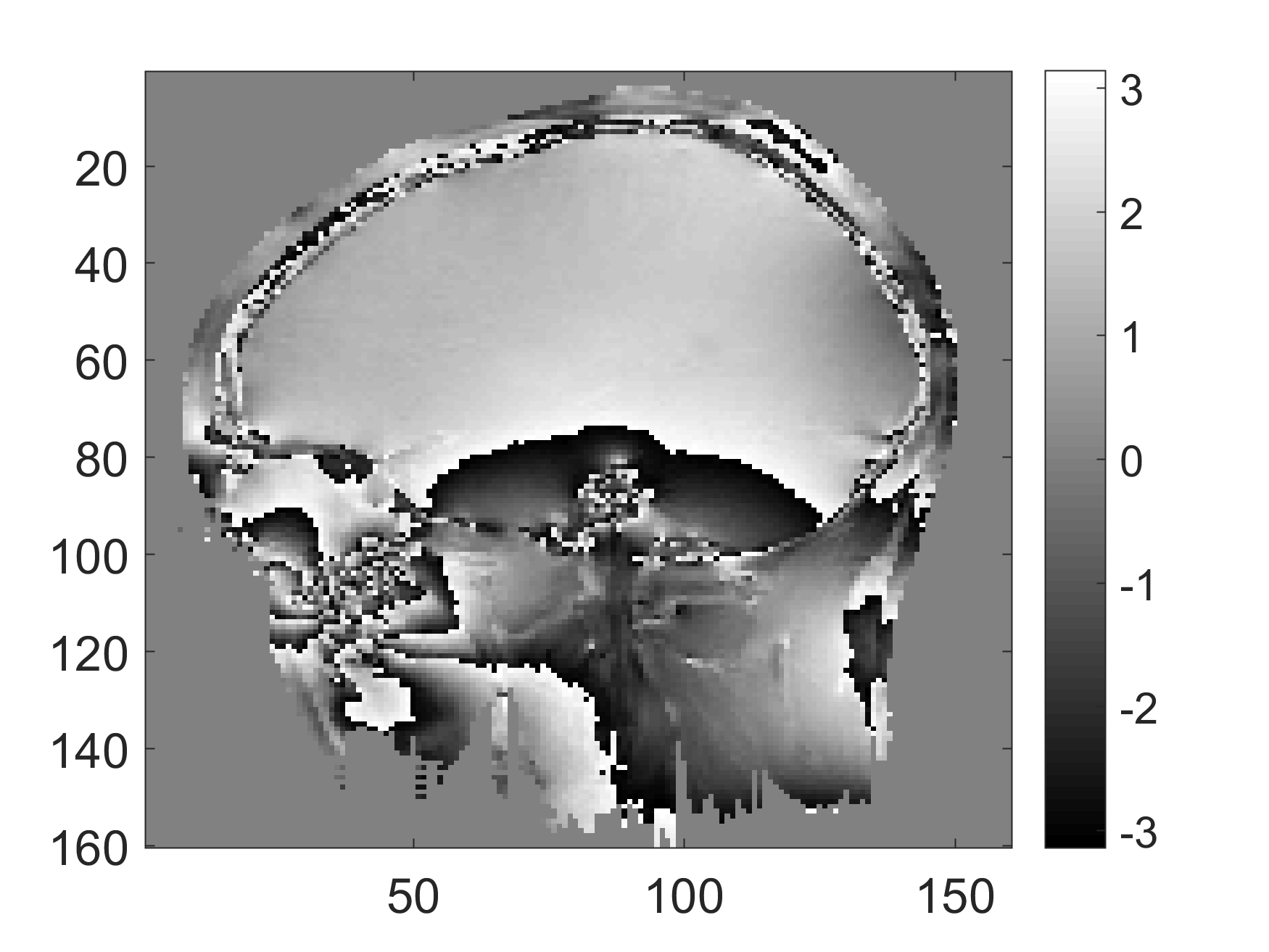}
		\caption{ side view}
		\label{mri_a1}
	\end{subfigure}
	\hfill
	\begin{subfigure}{0.32\textwidth}
		\includegraphics[width=\textwidth]{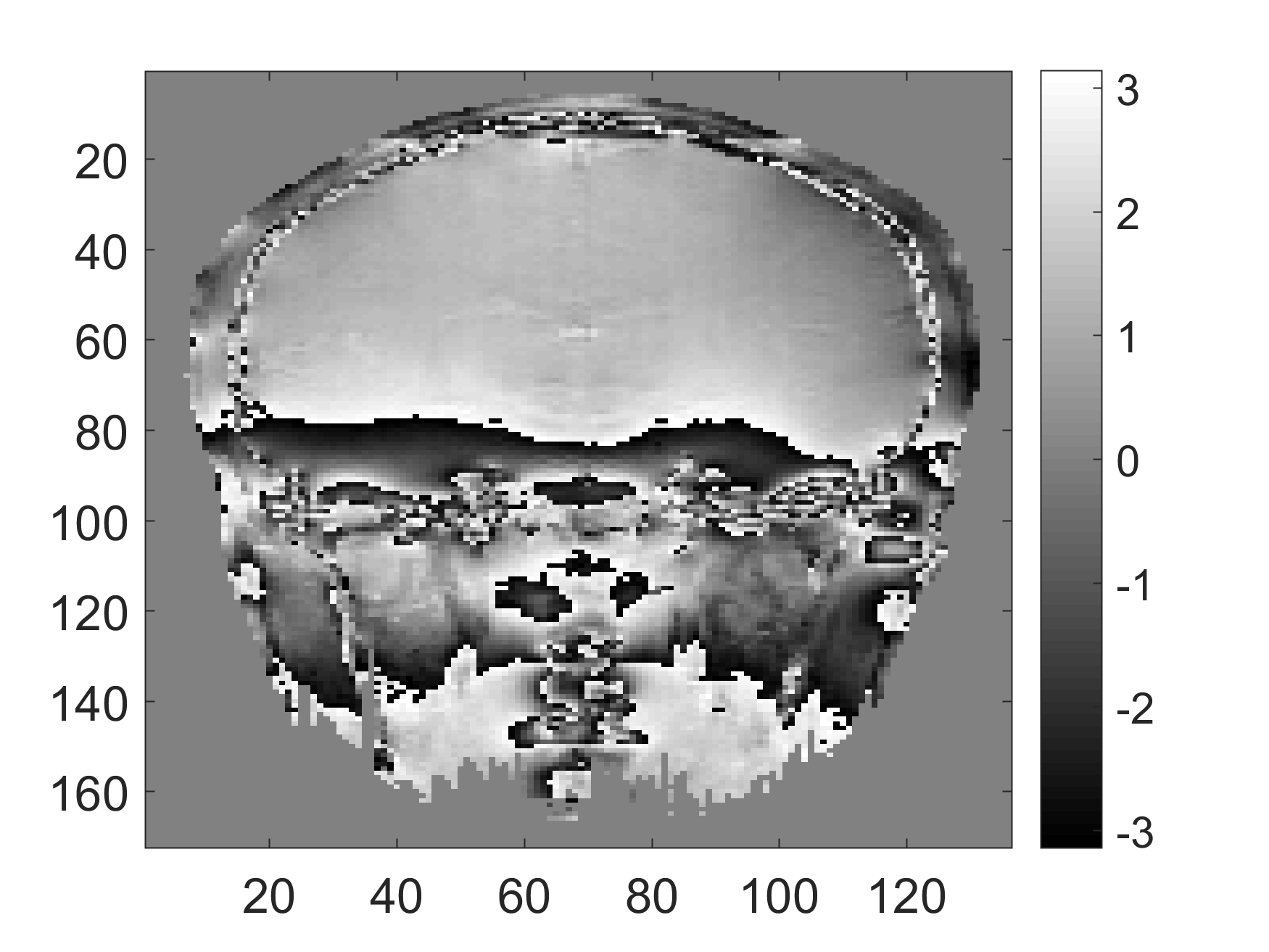}
		\caption{ front view}
		\label{mri_a2}
	\end{subfigure}
	\hfill
	\begin{subfigure}{0.32\textwidth}
		\includegraphics[width=\textwidth]{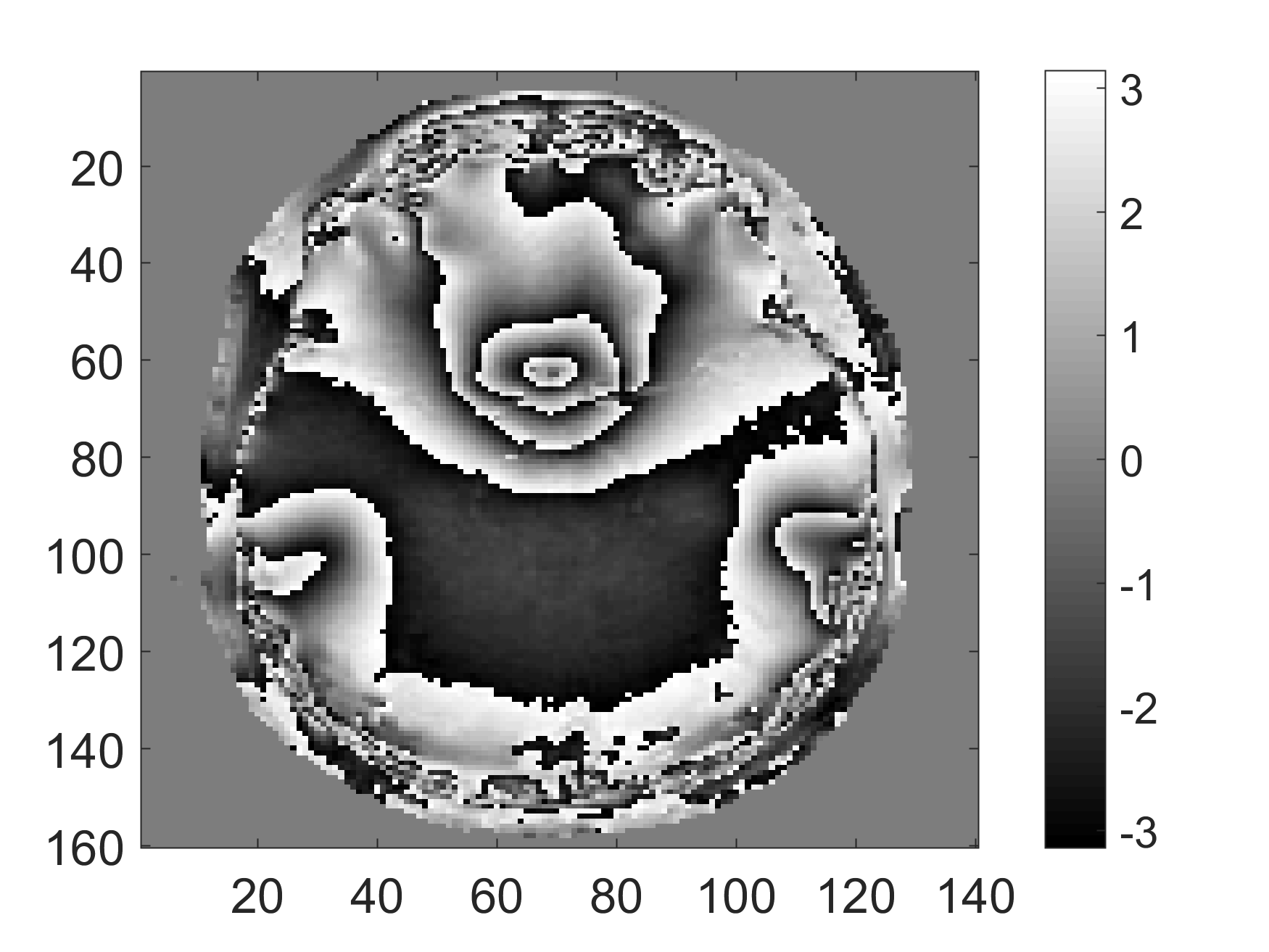}
		\caption{ top view}
		\label{mri_a3}
	\end{subfigure}
	\caption{ MRI interferograms along different orientations.}
	\label{mri_a}
\end{figure}

\begin{figure}[h!]
	\centering
	\begin{subfigure}{0.32\textwidth}
		\includegraphics[width=\textwidth,height=1.2\textwidth]{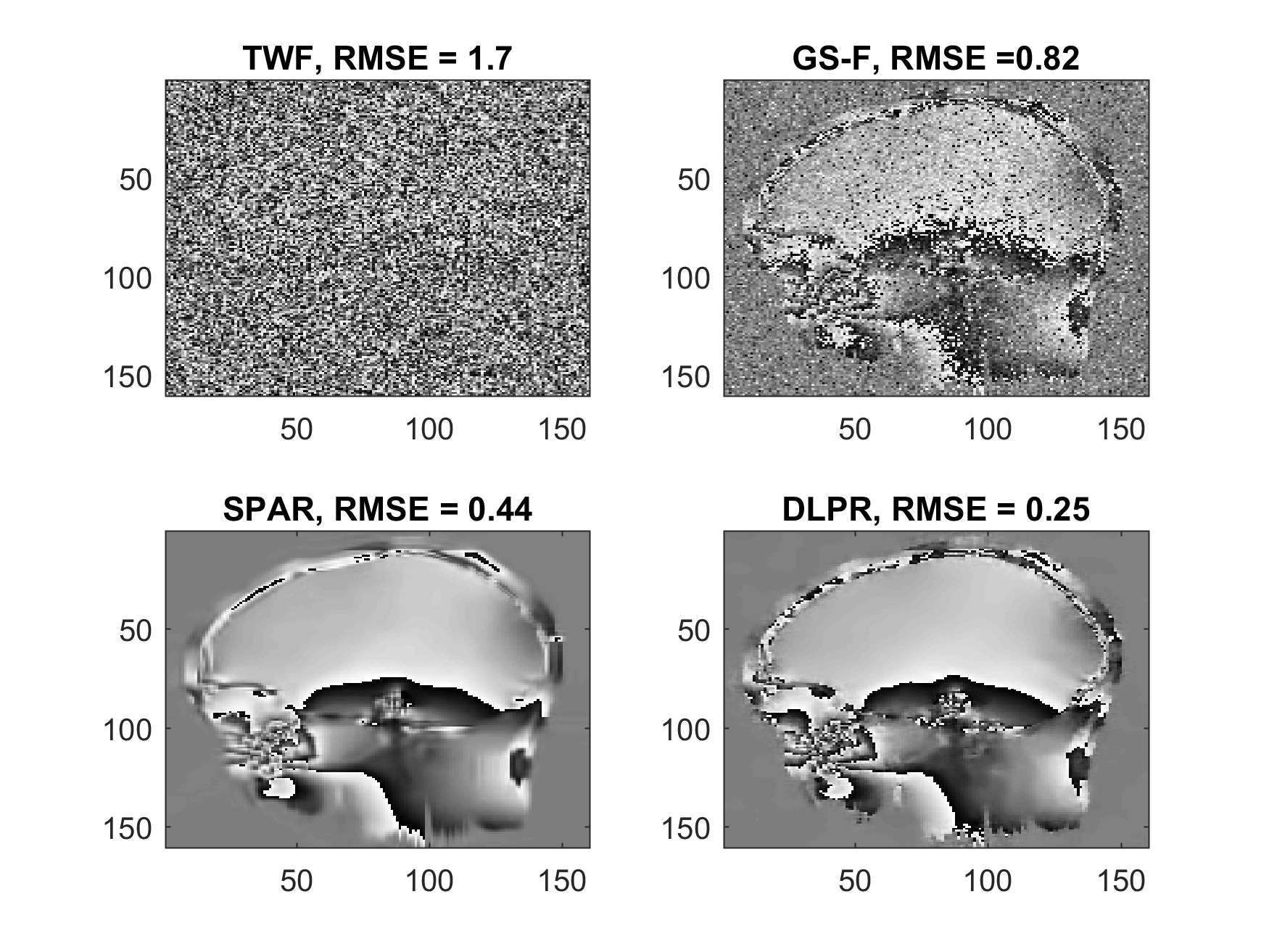}
		\caption{ side view}
		\label{mri_e1}
	\end{subfigure}
	\hfill
	\begin{subfigure}{0.32\textwidth}
		\includegraphics[width=\textwidth,height=1.2\textwidth]{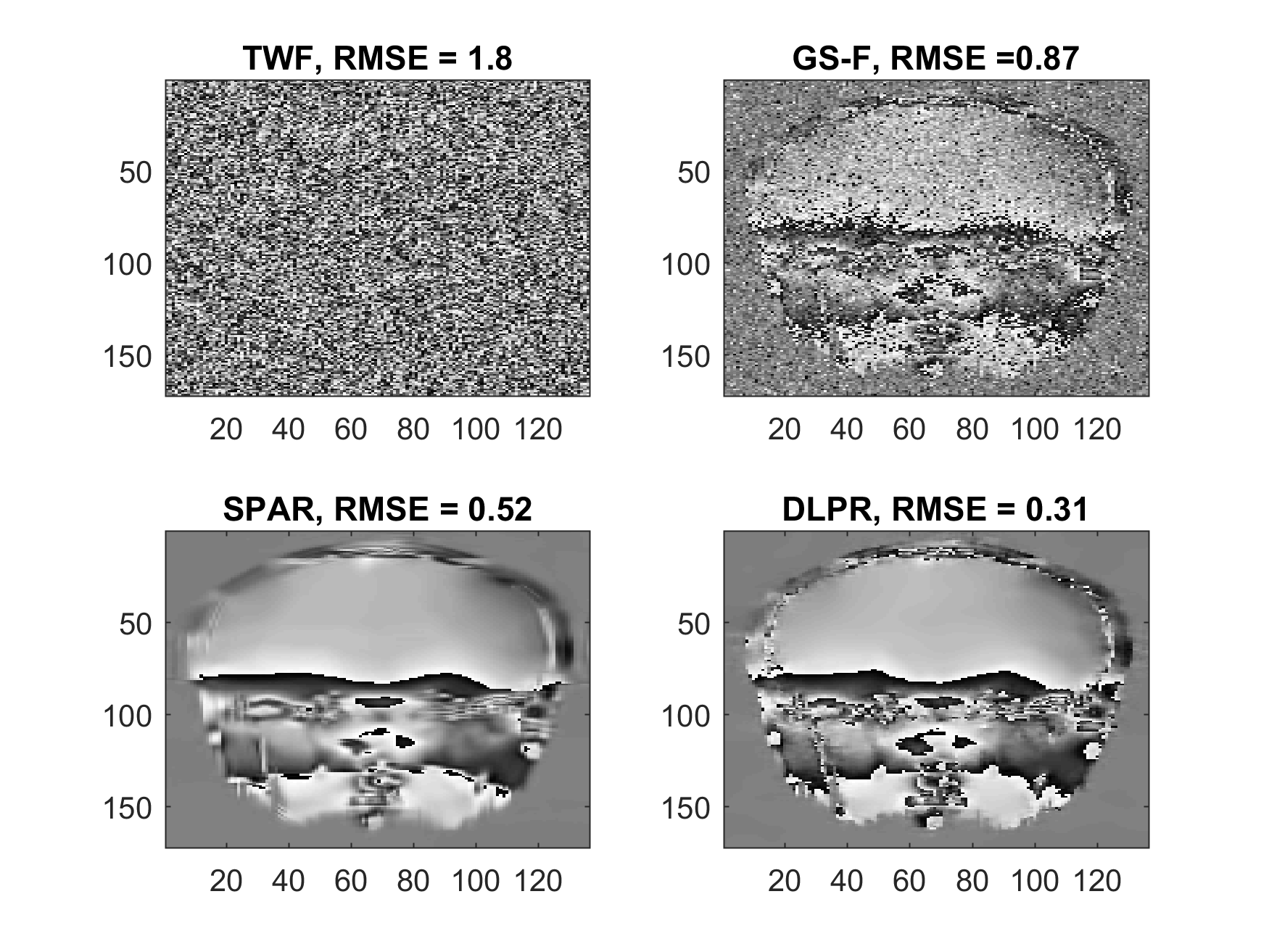}
		\caption{ front view}
		\label{mri_e2}
	\end{subfigure}
	\hfill
	\begin{subfigure}{0.32\textwidth}
		\includegraphics[width=\textwidth,height=1.2\textwidth]{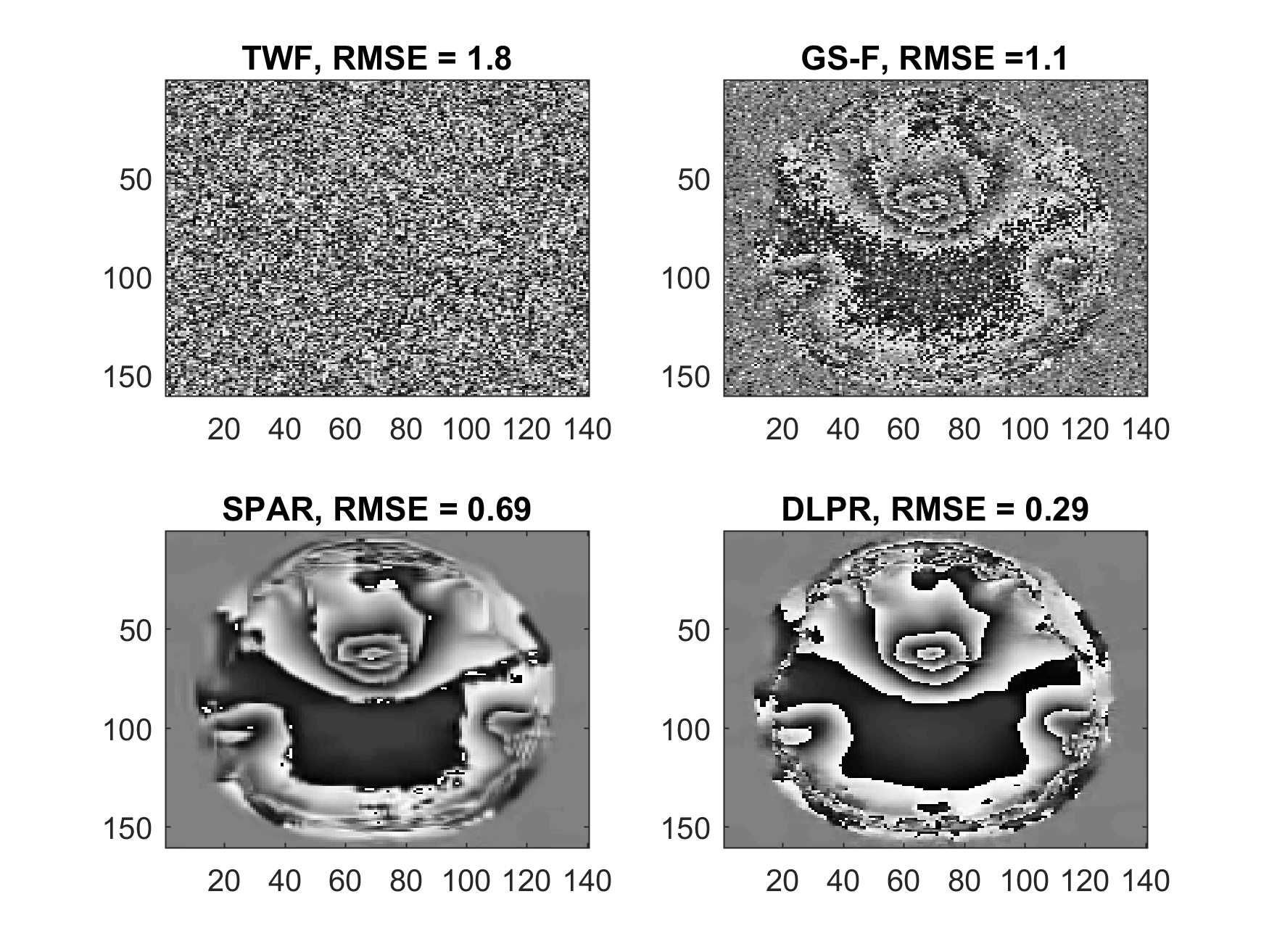}
		\caption{ top view}
		\label{mri_e3}
	\end{subfigure}
	\caption{ \small MRI interferograms retrieved using different PR algorithms. Poissonian observations with high noise level ($\chi=0.00001,~{\rm SNR}=-7~{\rm dB}$) is considered}
	\label{mri_e}
\end{figure}

\begin{figure}[h!]
	\centering
	\begin{subfigure}{0.32\textwidth}
		\includegraphics[width=\textwidth]{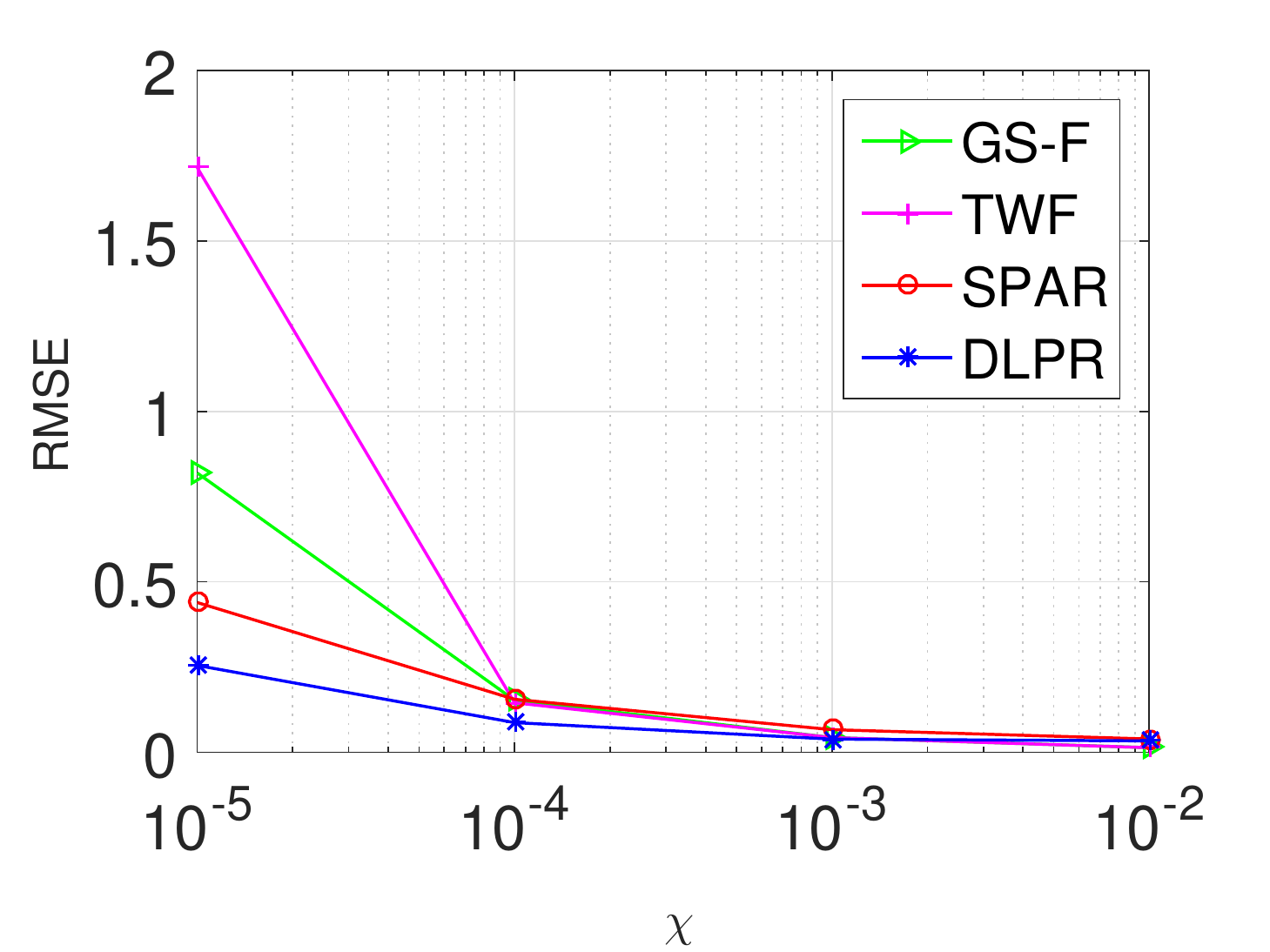}
		\caption{side view}
		\label{mri_b1}
	\end{subfigure}
	\hfill
	\begin{subfigure}{0.32\textwidth}
		\includegraphics[width=\textwidth]{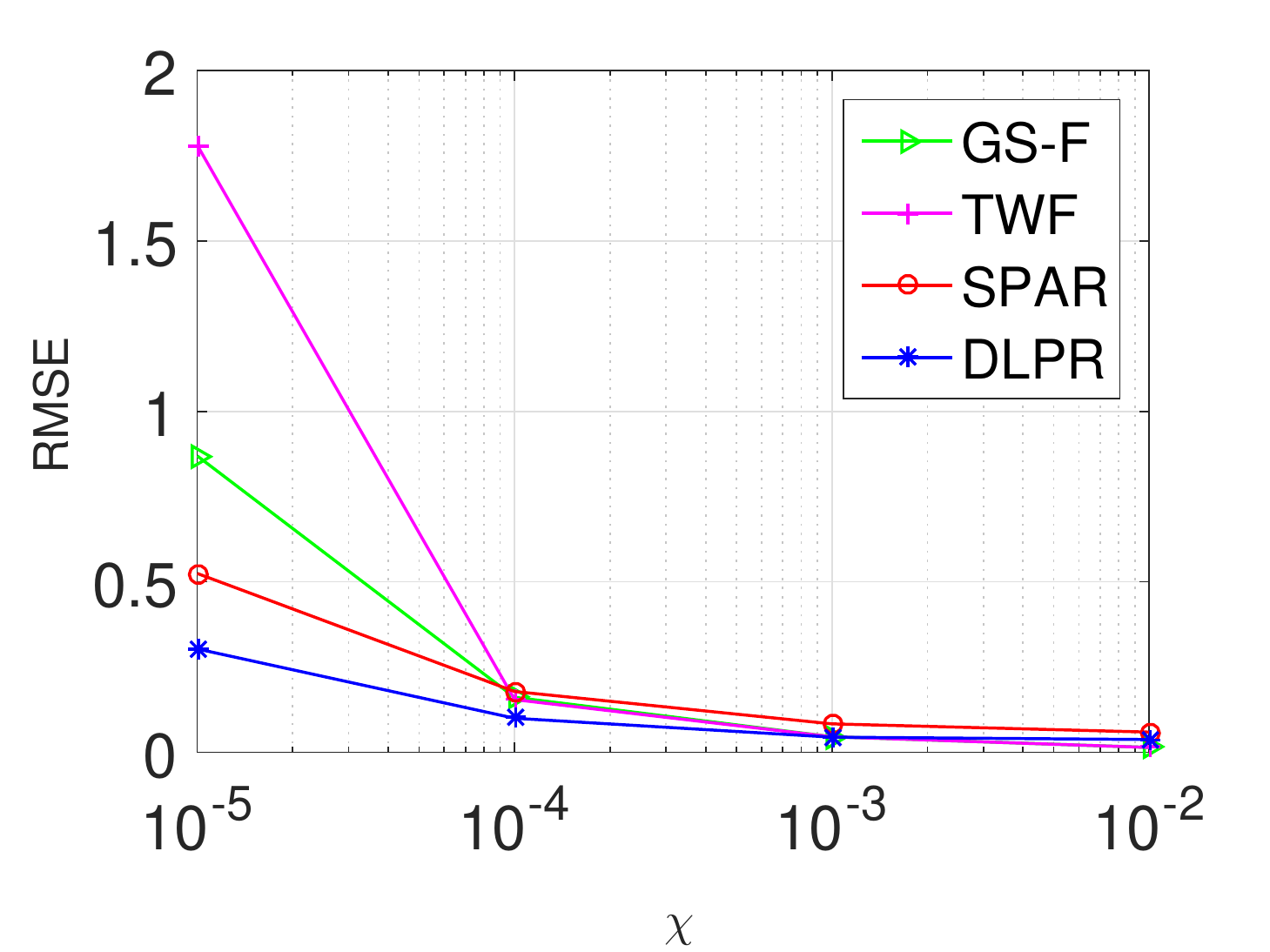}
		\caption{front view}
		\label{mri_b2}
	\end{subfigure}
	\hfill
	\begin{subfigure}{0.32\textwidth}
		\includegraphics[width=\textwidth]{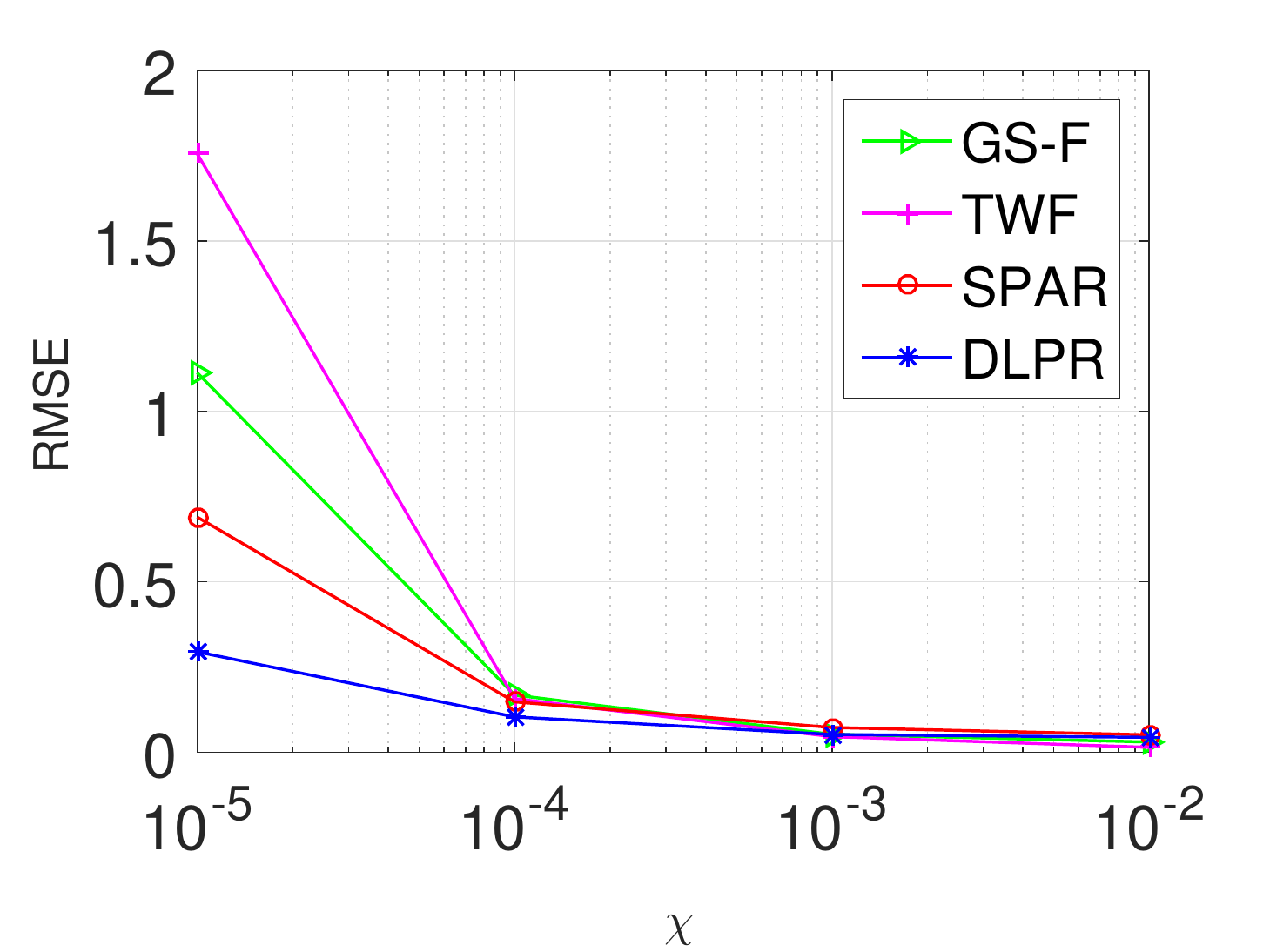}
		\caption{top view}
		\label{mri_b3}
	\end{subfigure}
	\caption{RMSE for MRI interferogram reconstruction. Poissonian observation model is considered.}
	\label{mri_b}
\end{figure}

Experiments using real MRI interferograms are included in this section. The MRI interferograms \footnote{The work was carried out on a 1.5 T GE Signa clinical scanner operating within Western General Hospital (WGH), University of Edinburgh.} used in this section are obtained by scanning the human head region along side, top and front orientations as shown in Fig. \ref{mri_a}. The complex-valued test images are generated by using each of these interferograms (figs. \ref{mri_a1}, \ref{mri_a2}, \ref{mri_a3}) as the clean phase ($\bs{\psi}_{2\pi}$) with amplitude set to unity. These complex-valued images are self-similar, and thus its patches are well approximated by sparse representations over a learned dictionary. Hence we emphasize that although an optical set-up is being considered in our discussion, these MRI interferograms are relevant test images. In comparison to the synthetic data used in the  previous experiments, these real interferograms do not have a well defined smooth structure and are highly challenging data for a PR experiment. Despite these challenges, DLPR remains to be a strong candidate proving its ability to retrieve real interferograms. The estimates for a highly noisy observation ($\chi=0.00001$) are shown in Fig. \ref{mri_e}. It can be qualitatively examined that DLPR estimates are much better in preserving the sharp details of the interferograms. This is clearly supported by the RMSE values. The results for other noise levels are shown in Fig. \ref{mri_b}, in the form of a graph, from which we can arrive at the same conclusion. These experiments show the strong ability of DLPR to retrieve real phase data.


\subsection{Gaussian Observation}
\begin{figure}[h!]
	\centering
	\begin{subfigure}{0.32\textwidth}
		\includegraphics[width=\textwidth]{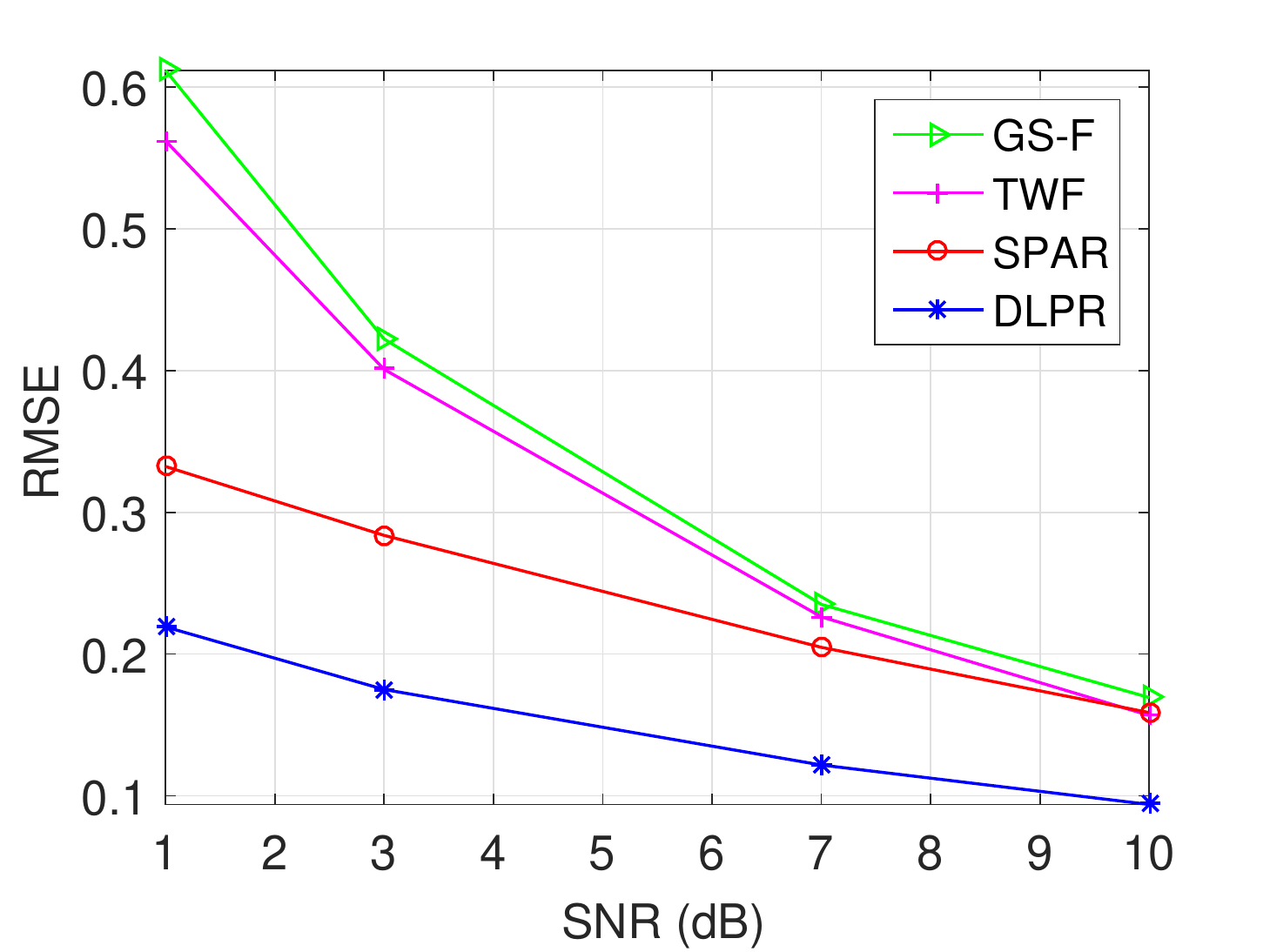}
		\caption{side view}
		\label{mri_ga1}
	\end{subfigure}
	\hfill
	\begin{subfigure}{0.32\textwidth}
		\includegraphics[width=\textwidth]{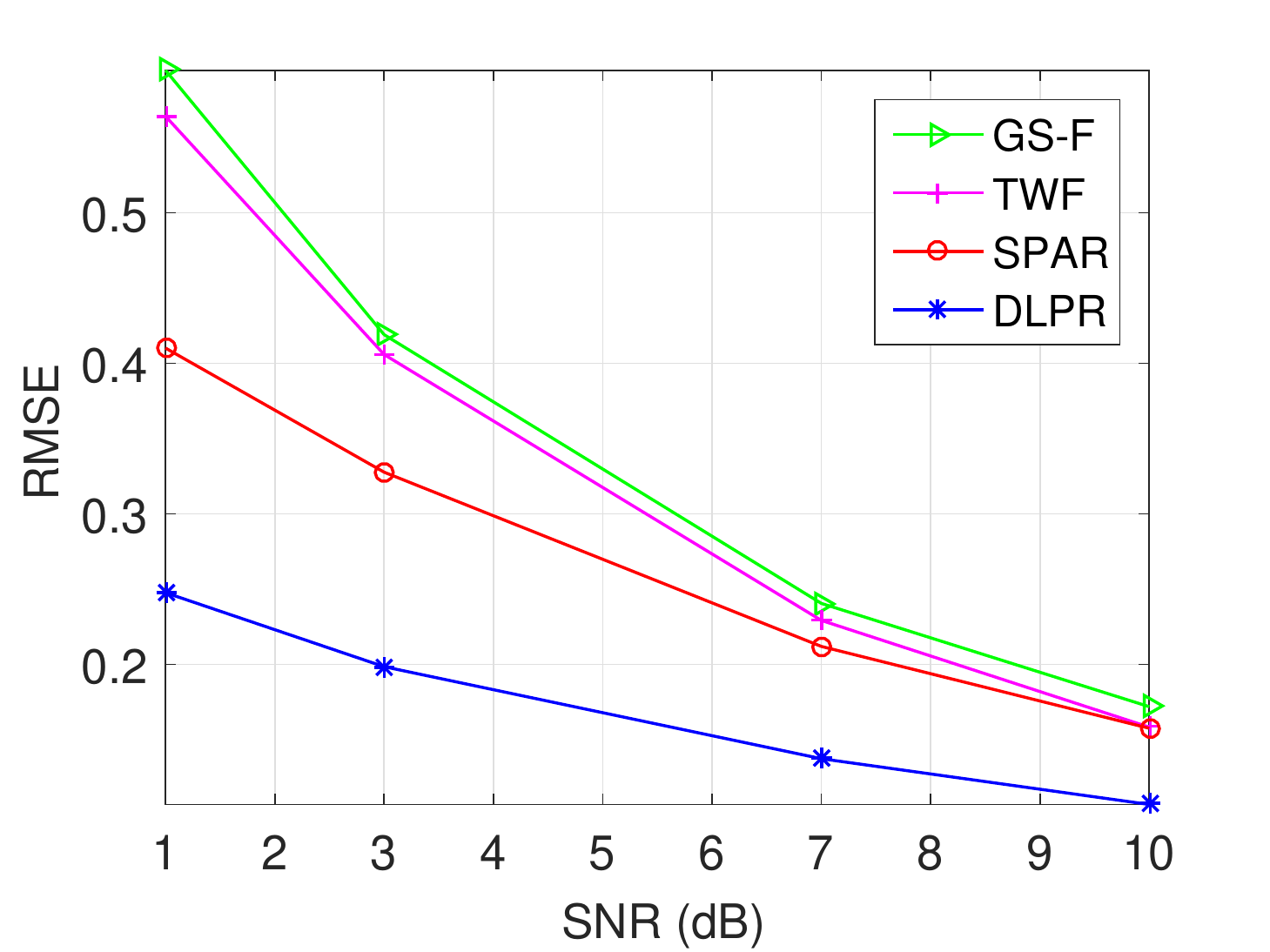}
		\caption{front view}
		\label{mri_ga2}
	\end{subfigure}
	\hfill
	\begin{subfigure}{0.32\textwidth}
		\includegraphics[width=\textwidth]{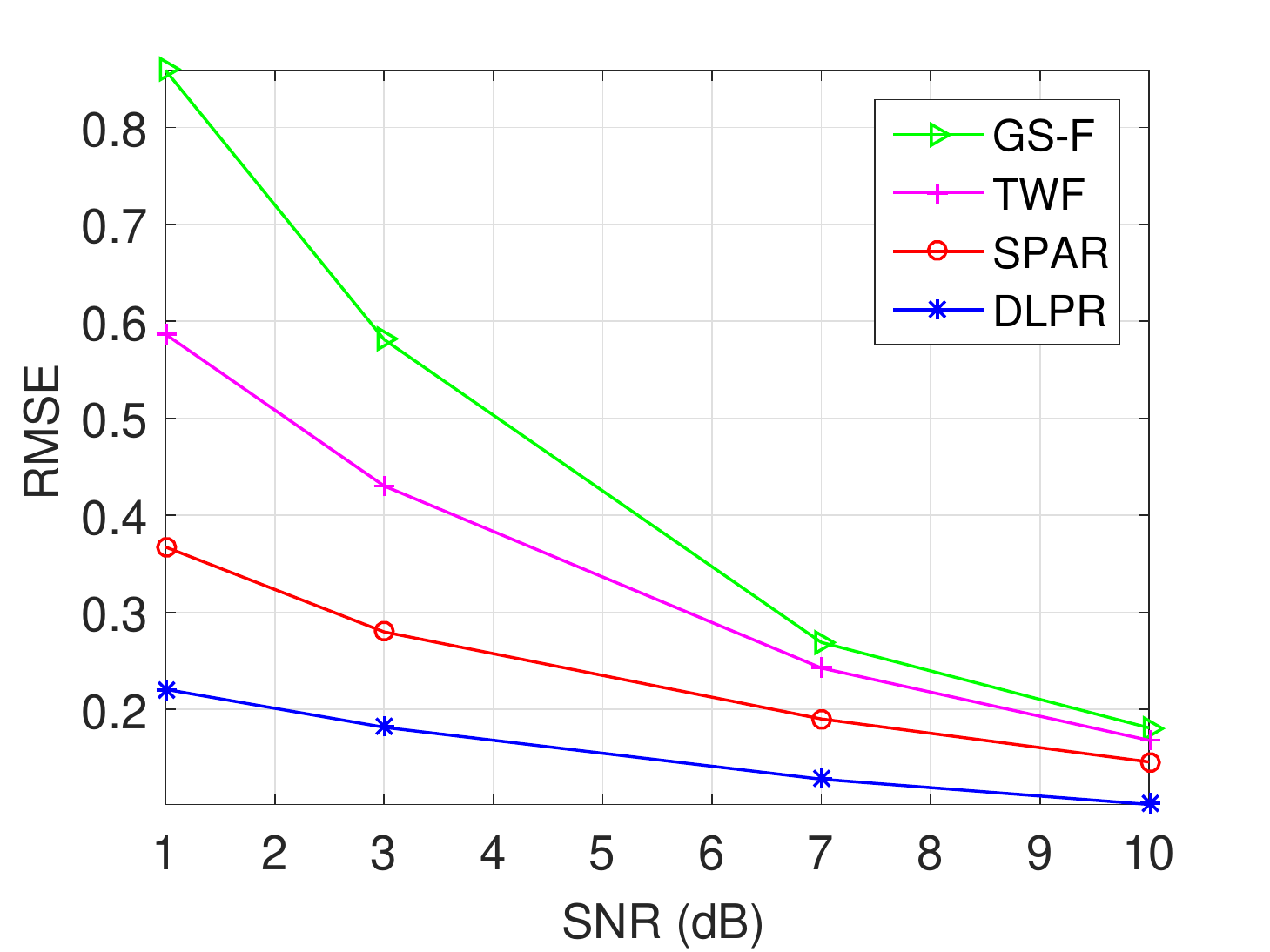}
		\caption{top view}
		\label{mri_ga3}
	\end{subfigure}
	\caption{RMSE for MRI interferogram reconstruction. Gaussian observation model is considered.}
	\label{mri_ga}
\end{figure}

In this section, PR is conducted by considering a Gaussian observation model. Very low SNR values, i.e.,  $\text{SNR} \in \cbr{1, 3, 7, 10}\text{dB} $ is considered corresponding to highly to moderate noisy scenarios. In this section, we are not repeating all the experiments conducted for Possonian case. But the toughest experiments with real MRI interferograms are repeated. The results are shown in Fig. \ref{mri_ga}, which indicate that DLPR performs exceptionally well and beats it competitors with a good margin for all noise levels.
\subsection{Prior-plugged DLPR for Class-specific Phase retrieval}
DLPR algorithm \ref{DLPRalg} learns the dictionary iteratively and retrieves the phase from observed intensities. But in many practical applications, the phase to be recovered is known to belong to certain classes of images. In such applications, the clean images from the specific classes can be used to learn prior information and to enhance the PR process. Most of the PR methods that we discussed does not have the algorithmic structure to incorporate such priors. But DLPR, being a dictionary based method, opens the door to the exploitation of the learned prior. Algorithm \ref{DLPRalgCspec}, which we term as \textit{prior-plugged} DLPR, is a modified version of the original DLPR Algorithm \ref{DLPRalg}, obtained by skipping the online dictionary learning step and including the learned dictionary $\bf D^{l}$ (prior) as input.

\begin{figure}[h!]
	\centering
	\begin{subfigure}{0.70\textwidth}
		\includegraphics[width=\textwidth]{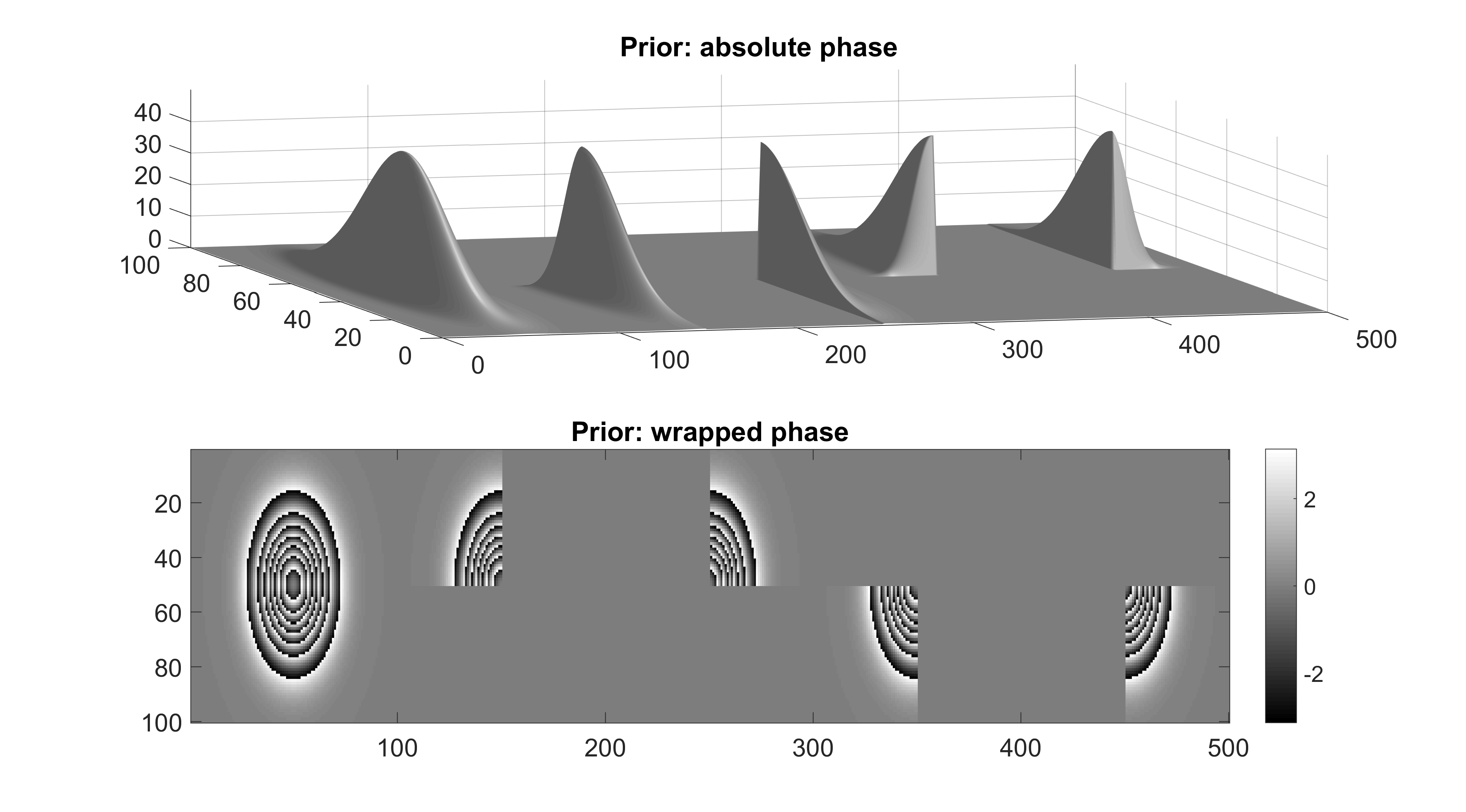}
		\caption{Gaussian and its truncated parts used to learn dictionary (prior)}
		\label{TruncatedGaussPrior}
	\end{subfigure}
	\hfill
	\begin{subfigure}{0.26\textwidth}
		\includegraphics[width=\textwidth]{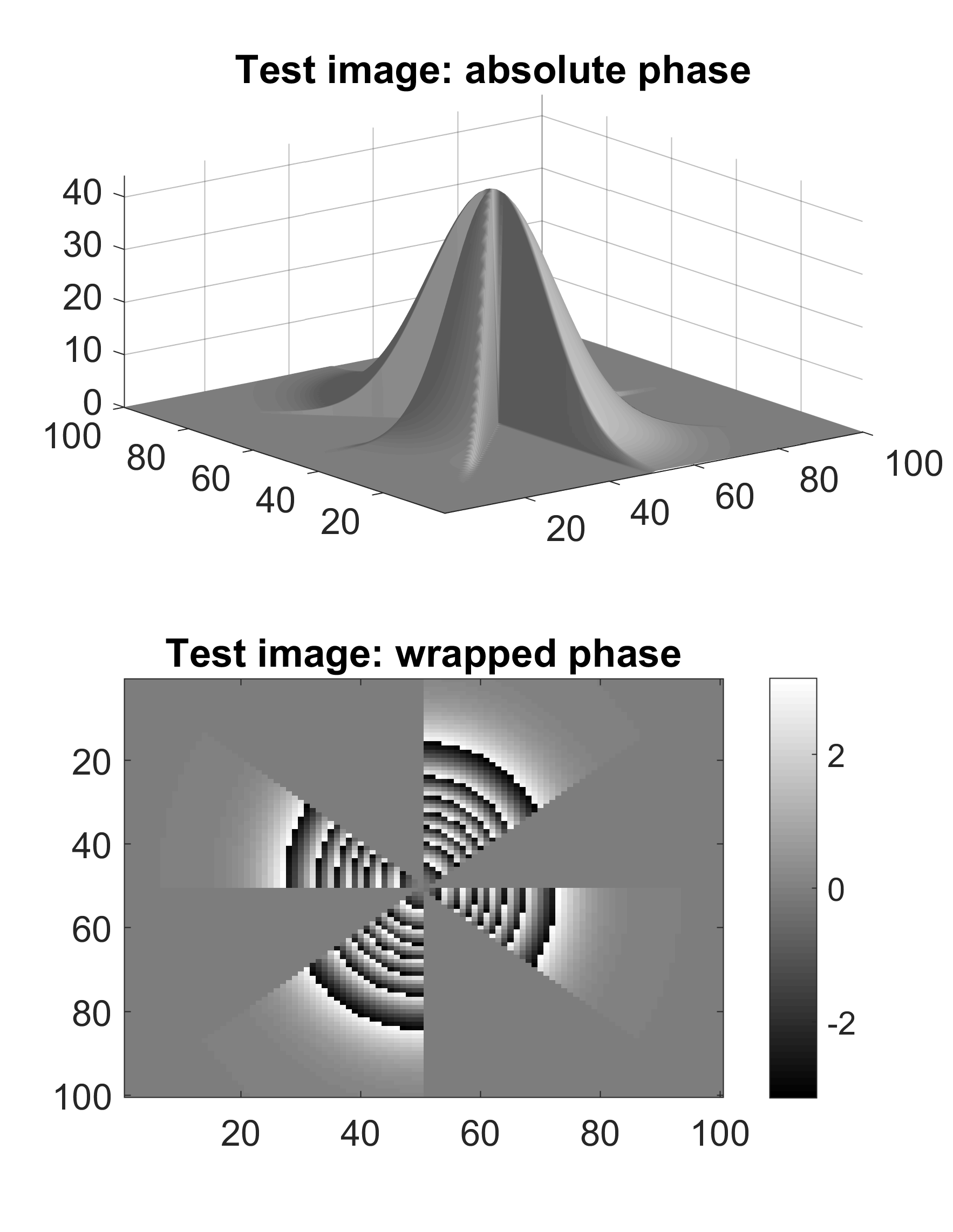}
		\caption{Test surface }
		\label{cake}
	\end{subfigure}
	\caption{Simulated data set for class-specific phase retrieval}
	\label{csTG}
\end{figure}

\begin{algorithm}[h!]
	\SetKwInOut{Input}{Input}\SetKwInOut{Output}{Output}\SetKwInOut{Init}{Init}
	\caption{Prior-plugged DLPR for class-specific PR}
	\label{DLPRalgCspec}
	\Input{$\mathbf{z}_{s}\in \mathbb{R}^{n}$ (noisy observation)\\
		\Init{$\widehat{\mathbf{x}}^0 \in \mathbb{C}^{n}$ (object wavefront), \\ $\mathbf{D}^{{l}}\in \mathbb{C}^{w^2\times k}$ (learned dictionary), \\ $\bs{\alpha}^{0}\in \mathbb{C}^{k}$ (code), \\$\gamma,\beta>0$(regularization  parameter), \\$S\in\mathbb{N}$ (Number of observations), \\ $T\in\mathbb{N}$ (number of iteration)}}
	\Output{$\widehat{\mathbf{x}}^T, \bs{\alpha}^T$} 
	\For{$t\leq T$}{
		\textbf{Forward Propagation:} \hspace{10cm}
		$\mathbf{\hat{v}}_{s}^{t}=\mathbf{A}_{s}\mathbf{\hat{x}}^{t-1}\text{, where } \mathbf{A}_{s}=\mathbf{FM}_{s}$ and $s=1,...,S$\\
		\textbf{Filtering at Sensor Plane:} \hspace{10cm} 
		$\mathbf{\hat{u}}_{s}^{t}=\mathbf{\hat{b}}_{s}^{t}\mathbf{\odot }\exp \nbr{j\cdot \text{angle}\nbr{\mathbf{\hat{v}}_{s}^{t}}}, s=1,...,S, $ where $\mathbf{\hat{b}}_{s}^{t}$ is given by \eqref{bslPoi} and \eqref{bslGau} for Poissonian and Gaussian observations respectively\\
		\textbf{Backward Propagation:} \hspace{2cm}
		$\bf{\hat{x}}^{t-1/2}=  \left(\sum_{s=1}^{S}\frac{\bf{A}_{s}^{H} \bf{A}_{s}}{\beta\gamma}+\sum_{i\in{\cal I}_p}\bf{R}_{i}^{H} \bf{R}_{i}\right)^{-1} \times  \left(\sum_{s=1}^{S} \frac{\bf{A}_{s}^{H}\bf{u}_{s}}{\beta\gamma}+\sum_{i\in{\cal I}_p}\bf{R}_{i}^{H} \bf{D}^{l}{\bs \alpha }_i^{t-1}\right)$, given by \eqref{bwdprop}\\
		\textbf{Patch formation:}  \hspace{10cm}
		$\bf{\hat{x}}_i^{p, t-1/2}= {\bf R}_i\bf{\hat{x}}^{t-1/2},~\forall i\in {\cal I}_p$ \\
		\textbf{Sparse Code learning:}  \hspace{10cm}
		Obtain $\bs{\alpha}_i^t$ by running OMP (Algorithm \ref{algOMP}) iterations  using $\bf{\hat{x}}_i^{p, t-1/2}$ and $\bf{D}^l$ as input\\
		\textbf{Sparse wavefront modeling at the object plane:}  \hspace{10cm}
		$\widehat{\bf{x}}_i^{p,t}=\bf{D}^l{\bs {\alpha}}_i^t,~\forall i\in {\cal I}_p$\\
		\textbf{Patch to image formation:}  \hspace{4cm}
		$\widehat{\bf{x}}^t=\nbr{\sum_{i\in{\cal I}_p}\bf{R}_{i}^{H} \bf{R}_{i}}^{-1}\nbr{\sum_{i\in{\cal I}_p}\bf{R}_{i}^{H}\widehat{\bf{x}}_i^{p,t}} $ from\eqref{p2imag}
	}
\end{algorithm}
In this section, the class-specific phase retrieval using prior-plugged DLPR is illustrated using two experiments, the first one using simulated data and the second one using real data. In the first experiment, to learn the prior (dictionary), a Gaussian surface and its four quarters, as shown in Fig. \ref{TruncatedGaussPrior}, is considered. The C-ODL Algorithm \ref{algCodl} is used to learn the dictionary. PR is performed on a truncated Gaussian surface created by cutting off the alternate octants of a Gaussian surface, as shown in Fig. \ref{cake}. We remark that the retrieval of this surface is quite challenging as it contains less connected areas and has a lot of sharp discontinuities. Though the prior and the testing surface are different, they belong to the same class of truncated Gaussians and share many identical regions. 

\begin{figure}[h!]
	\centering
	\begin{subfigure}{0.49\textwidth}
		\includegraphics[width=\textwidth,height=5.5cm]{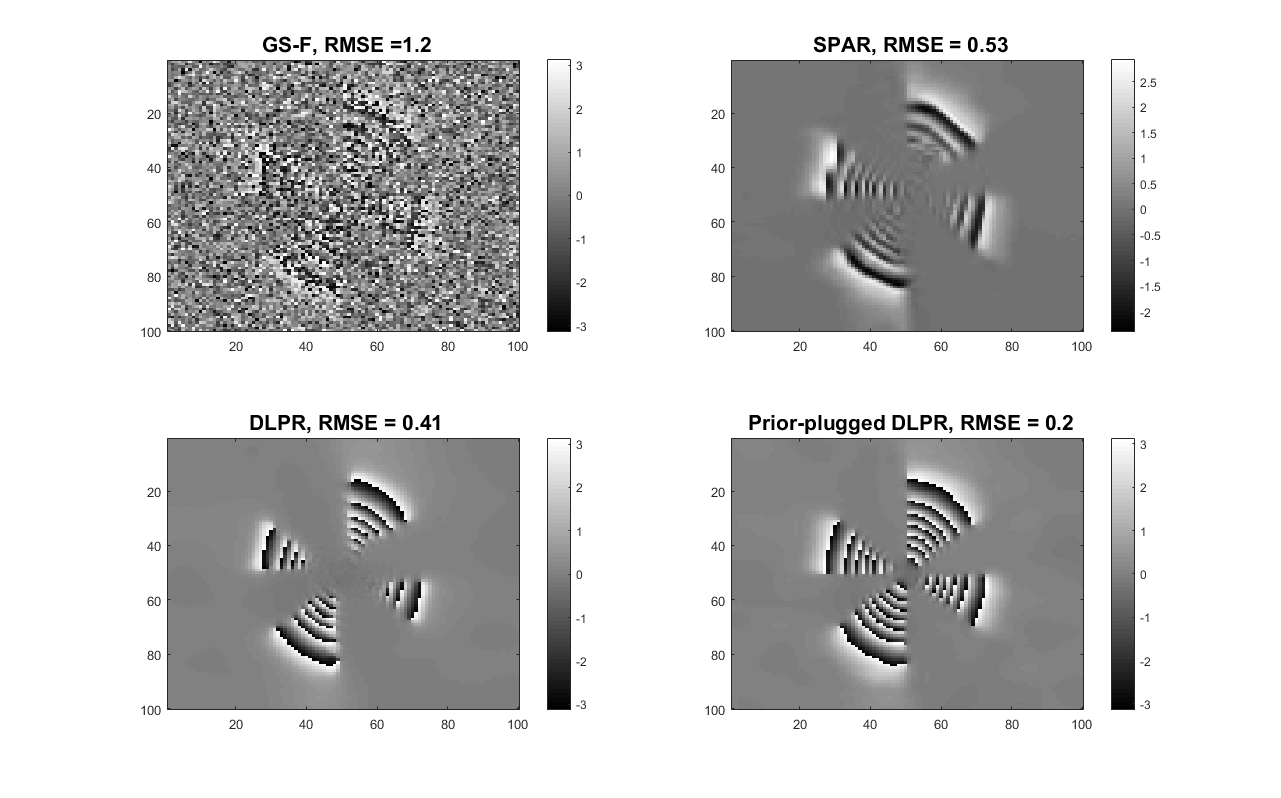} 	
		\caption{  Retrieved Phase using various methods. Poissonian observations ($\chi=0.00001,~{\rm SNR}=-7~{\rm dB}$) is used.}
		\label{TG_prior_Results}
	\end{subfigure}
	\hfill
	\begin{subfigure}{0.49\textwidth}
		\includegraphics[width=\textwidth,height=5.5cm]{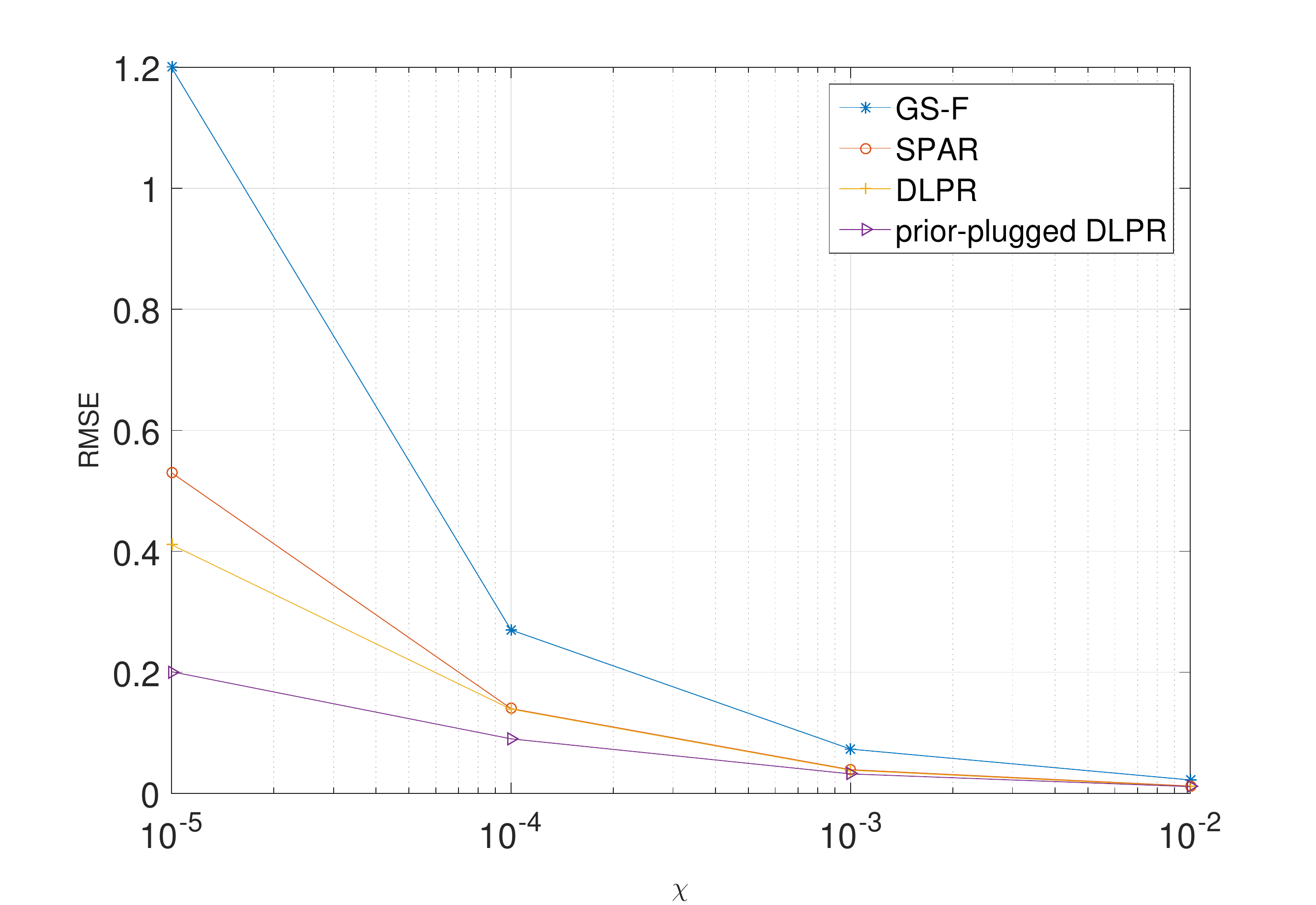} 	
		\caption{ Prior plugged DLPR versus GS-F, SPAR, DLPR for  Poissonian observations.}
		\label{TG_pr_RMSE}
	\end{subfigure}
	\caption{Performance of prior-plugged DLPR for the experiments conducted using phase data belonging to truncated Gaussian family}
	\label{}
\end{figure}

Fig. \ref{TG_prior_Results} shows the results for prior-plugged DLPR in comparison with GS-F, SPAR, and DLPR using very high Poissonian noise ($\chi=0.00001$). It is quite evident that the prior-plugged DLPR shows remarkable improvement by exploiting the learned prior. In comparison with the original DLPR, which does not use the prior, the prior-plugged DLPR has an improvement in RMSE by $0.41-0.20=0.21$. The results for similar experiments by considering other noise levels are plotted in Fig. \ref{TG_pr_RMSE} in which prior-plugged DLPR beats all other methods, especially for the highly noisy cases.  

\begin{figure}[h!]
	\centering
	\includegraphics[scale=0.5]{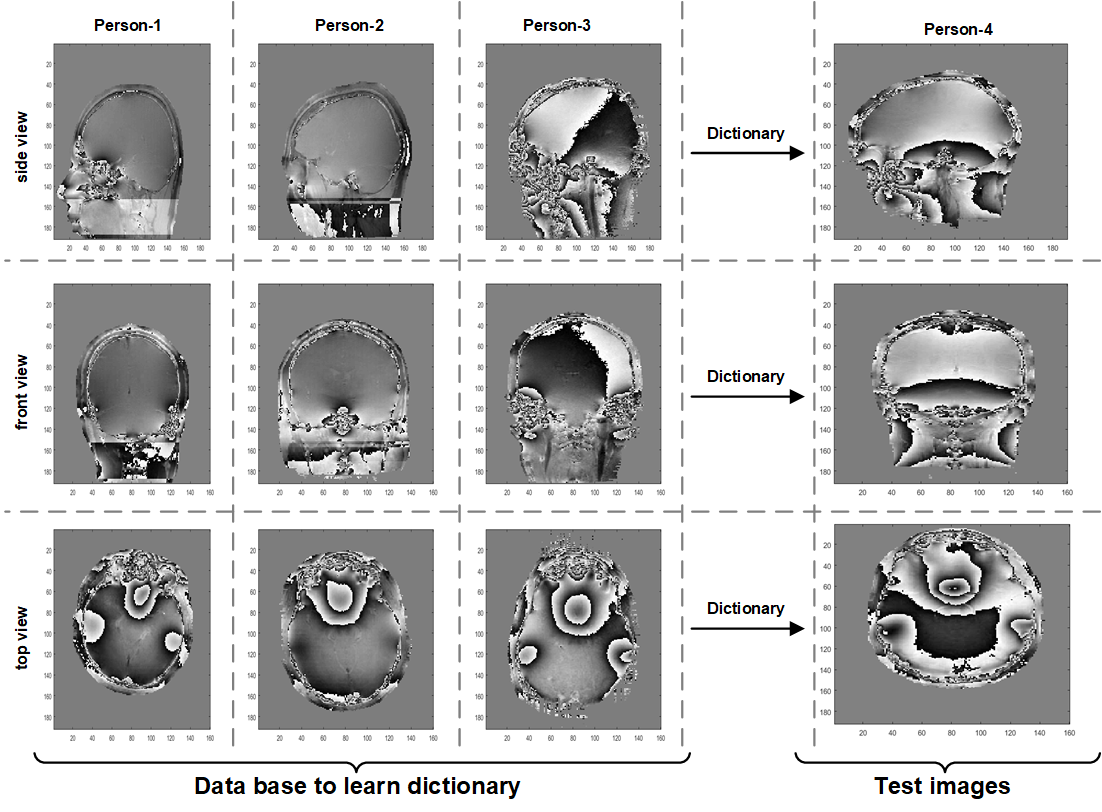} 	
	\caption{Real MRI data set for class specific phase retrieval.}
	\label{csmri}
\end{figure}

The remarkable improvement in RMSE in the above experiment is due to the fact that both the prior-learning surface and the phase to be retrieved share many identical regions. But in practice this may not be true and we only have access to data from the same class. The phase to be retrieved need not possess regions exactly identical with the training images. To account this factor, we present a second experiment using real MRI data. The details of the data set \footnote[3]{The work was carried out on a 1.5 T GE Signa clinical scanner operating within Western General Hospital (WGH), University of Edinburgh.} is shown in Fig. \ref{csmri}. 

In this experiment, the MRI phase images obtained from four different persons are used. These images are the interferograms of the head region obtained through the scanning along front, side and top orientations. In the following experiment, a particular scanning orientation, say front view, is taken as a specific class. The dictionary is learned from the front view interferograms of person-1,2 and 3. This dictionary is used as the prior for retrieving the interferograms of the fourth person along front view. In the PR experiment, a very high Poissonian noise corresponding to $\chi=0.00001$ (${\rm SNR}=-7~{\rm dB}$) is considered. The experiment is repeated for other two scanning orientations, i.e., side view and top view. The result in terms of RMSE is given in Table \ref{mri_table}. From the table, it is evident that prior-plugged DLPR is much better compared to SPAR, GS-F, and TWF and yields slightly better results compared to normal DLPR. We emphasis that in prior-plugged DLPR, the fourth person's MRI phase image is retrieved just by using the dictionary learned from the scan images of the other three persons and still, the obtained results are better than the normal DLPR. This proves that prior-plugged DLPR is a powerful tool for class-specific phase retrieval applications.
\begin{table}[h!]
	\begin{center}
		\caption{Performance evaluation for class-specific phase retrieval using real MRI data. The best values are shown in bold font.}
		\label{mri_table}
		\begin{tabular}{r|ccccc}
			\multicolumn{1}{c|}{\multirow{2}[1]{*}{}}  & \multicolumn{4}{c}{ \B{RMSE}} &\\
			\multicolumn{1}{c|}{\multirow{2}[1]{*}{Surf.}}  & \multicolumn{4}{c}{} &prior-plugged\\
			&      GS-F & TWF & SPAR & DLPR & DLPR  \\
			
			\midrule
			\midrule
			Side view           & 0.587	    &1.560  &0.330  &0.202 &\B{0.184}\\
			\midrule
			\midrule
			Top view            &0.707	    &1.789	&0.366  &0.220 &\B{0.194}\\
			\midrule
			\midrule
			Front view          &0.698  	    &1.693	&0.393 &0.226  &\B{0.192}   \\
			\midrule
			\midrule
		\end{tabular}
	\end{center}
\end{table}

\subsection{Complexity of DLPR}
The complexity of DLPR is mainly defined by the dictionary learning and OMP steps. A theoretical analysis on this is beyond the scope of the work presented. However, we characterise the complexity by the time required to retrieve the phase. All the results presented in this paper are obtained through 20 iterations of DLPR and 50 iterations (as suggested by SPAR) of SPAR, GS-F, and TWF. We use MATLAB R2015b and computer with the processor Intel(R) Core(TM) i7-4790 CPU @3.60GHz, 16GB RAM. With the above mentioned number of iterations using an image of size $100 \times 100$ under Poissonian noise ($\chi=0.00001$), the time required for phase retrieval is as follows: DLPR$\simeq$ 77.6 sec, prior-pluged DLPR$\simeq$ 23.6 sec, SPAR$\simeq$ 15 sec, GS-F$\simeq$ 0.43 sec, TWF$\simeq$ 0.89 sec. It is to be noted that, in terms of time of convergence, DLPR lags behind SPAR due to its dictionary learning and sparse coding sub iterations that are computationally demanding.

\section{Conclusion}
Phase Retrieval from noisy intensity observations is addressed in this paper. A noisy optical system with wavefront modification is adopted for the discussion. The conventional PR algorithms fail to retrieve good quality phase images from noisy observations. To address this issue, a novel algorithm, termed as DLPR has been proposed in this paper. DLPR is an iterative algorithm developed under the alternating minimization framework and incorporates specially designed filters both at the sensor and the object plane of the optical imaging system. At the object plane, a sparse wavefront modeling using complex valued dictionary is done to obtain noise immunity. A second filtering mechanism is applied at the sensor plane aiming at additional noise suppression and is designed for Poissonian and for Gaussian observations. All the filters and optic propagation models are designed in a variational framework that maximizes the likelihood function and is aimed to iterate towards  statistically optimal estimates. 

The performance of the proposed algorithm is tested using challenging simulated and real data set. The state-of-the-art PR algorithms, namely SPAR, TWF, and GS-F, are used for performance comparisons. DLPR shows remarkable improvements over the conventional GS family and the recent TWF algorithm for highly noisy observations and its performance is quite competitive to  BM3D-based SPAR. It is shown that DLPR beats SPAR for highly noisy observations. DLPR also has an added advantage of prior exploitation in class-specific phase retrieval over SPAR, which was empirically demonstrated. As a future work, we suggest further researches on dictionary-based phase retrieval to improve the convergence speed of the algorithm by exploiting fast dictionary and sparse coding techniques.

\vspace{6pt} 
%
%
%
\authorcontributions{ Investigation, J.P. K.; Methodology, V.K; Supervision, J.M.B and V.K; Validation, J.P. K; Writing – original draft,  J.P. K ; Writing – review \& editing, J.M.B. }
%
\funding{The research leading to these results has received funding from the European Union’s H2020 Framework Programme (H2020-MSCA-ITN-2014) under grant agreement no 642685 MacSeNet. This work is also supported by the Portuguese Fundacão para a Ciência e Tecnologia (FCT) under grants UID/EEA/5008/2013 and Academy of Finland, project no. 287150, 2015-2019.}
%
\acknowledgments{The authors would like to thank Wajiha Bano, Institute for Digital Communications, University of Edinburgh, for providing the MRI data used in the experiment section of the manuscript.}
%
\conflictsofinterest{The authors declare no conflict of interest. The founding sponsors had no role in the design of the study; in the collection, analyses, or interpretation of data; in the writing of the manuscript, or in the decision to publish the results.} 

\reftitle{References}
\renewcommand\bf{\bfseries}

\bibliographystyle{IEEEbib}
\renewcommand\refname{References}
\bibliography{InPhaseRef}





\end{document}